\documentclass[12pt]{iopart}

\usepackage{graphicx}
\usepackage{amssymb}

\begin{document}

\title[Agent Based Models of Language Competition] {Agent Based Models
  of Language Competition: Macroscopic descriptions and Order-Disorder 
transitions}

\author{F. Vazquez, X. Castell\'o and M. San Miguel}

\address{IFISC, Institut de F\'isica Interdisciplin\`aria i Sistemes
  Complexos (CSIC-UIB), Campus Universitat Illes Balears, E-07122
  Palma de Mallorca, Spain}

\ead{federico@ifisc.uib-csic.es}

\begin{abstract} We investigate the dynamics of two agent based models of 
language competition.  In the first model, each individual can be in
one of two possible states,  either using language $X$ or  language
$Y$, while the second model  incorporates a third state $XY$,
representing individuals that use both  languages (bilinguals).   
We analyze the
models on complex networks and two-dimensional square lattices by
analytical and numerical methods, and show  that they exhibit a
transition from  one-language dominance to language coexistence.  We
find that the coexistence of languages is more difficult to  maintain
in the Bilinguals model, where the presence of bilinguals in use
facilitates the ultimate dominance of one of the two languages.  A
stability analysis reveals that the coexistence is more unlikely to
happen in poorly-connected than in fully connected networks, and that
the dominance of only one language is enhanced as the connectivity
decreases.  This dominance effect is even stronger in a
two-dimensional space, where domain coarsening  tends to drive the
system towards language consensus. 
\end{abstract}

\maketitle

\section{Introduction}
\label{intro}

A deep understanding of collective phenomena in Statistical Mechanics
is well established  in terms of microscopic spin models. Useful
macroscopic descriptions of these models in terms of mean field
approaches, pair and higher order approximations, and field theories
are also well known. Partly inspired by this success, collective
social phenomena are being currently studied in terms of microscopic
models of interacting agents  \cite{Castellano-09-a}. Agents, playing
here the role of spins, sit in the nodes of a network of social
interactions and change their state (social option) according to
specified dynamical rules of interaction with their neighbors in the
network. A general question addressed is the consensus problem,
reminiscent of order-disorder transitions: The aim is to establish
ranges of the parameters determining the interaction rules and network
characteristics for which the system is eventually dominated by one
state or option or, on the contrary, when a configuration of global
coexistence is reached \cite{SanMiguel-05_CISE}.

Language competition falls within the context of such social consensus
problems: One considers a population of agents that can use either of
two languages (two states). The agents change their state of using one
or the other language, by interactions with other agents. One is here
interested in determining when a state of dominance (or extinction) of
one language is reached, or when a state of global language
coexistence,  with a finite fraction of the two kind of speakers,
prevails. A particular and interesting ingredient in this problem is
the possibility of a third state associated with bilingual agents,
which have been claimed to play an essential role in processes of
language contact \cite{Appel_1987,Mira-05}. A good deal of work along
these lines originates in a paper by Abrams and Strogatz
\cite{Abrams-03}. These authors introduced a simple population
dynamics model with the aim of describing the irreversible death of
many languages around the world \cite{Crystal-00}. The original
Abrams-Strogatz model (ASM)   was a macroscopic description based on
ordinary differential equations, but a corresponding microscopic agent
based model was described in \cite{Stauffer-07}.  This model features
probabilities to switch languages determined by the local density of
speakers of the opposite language, and by two parameters that we call 
\emph{prestige}, $\mathcal S$ and \emph{volatility},
$a$. Prestige is a symmetry breaking parameter favoring the state
associated with one or other language which accounts for the
differences in the social status between the two languages in
competition. The volatility parameter determines the functional form
of the switching probability. It characterizes a property of the
social dynamics associated to the inertia of an agent to change its
current option, with its neutral value $a=1$ corresponding to a
mechanism of random imitation of a neighbor in the network.  Other
studies of the ASM  account for the effects of geographical boundaries
introduced in terms of reaction-diffusion equations
\cite{Patriarca_2008} or for Lotka-Volterra modifications of the
original model \cite{Pinasco-06}. Another different class of models
accounts for many languages in competition, with the aim of
reproducing the distribution of language sizes in the world
\cite{Schulze-05,Vivane-06(1)}.

While in the original paper of Abrams and Strogatz \cite{Abrams-03}
the two parameters $\mathcal S$ and $a$ were fitted to a particular
linguistic data, most subsequent work has focused on theoretical
analysis for the case of symmetrical prestige and neutral volatility.
For these parameter values (symmetric $\mathcal S$ and $a=1$) the
microscopic ASM coincides with the voter model \cite{Ligget_1985}, a
paradigmatic spin model of nonequilibrium dynamics
\cite{Marro_Dickman_1999}.  Inspired in the modifications proposed by
Minett and Wang of the ASM  \cite{Wang-05_TRENDS_Ecology,Minett-08}, a
microscopic Bilinguals model (BM) which introduces a third (intermediate)
state to account for bilingualism has been studied for the case of 
symmetric $\mathcal S$ and $a=1$ in \cite{Castello-06}.  In this way, 
this case is an extension of the original voter model.  The emphasis has 
been in describing the effects of the third state of bilingual agents in 
the dynamics of language competition as compared with the reference case
provided by the voter model. This includes the characterization of the
different processes of domain growth \cite{Castello-06}, and the role
of the network topology, like small world networks \cite{Castello-06}
and networks with mesoscopic community structure
\cite{Castello-07,Toivonen-08}.  Other studies associated to
variations of the voter model dynamics and the addition of intermediate
states have also been addressed in
\cite{Dall'Asta-07,Stark-08,blythe2009gmc,Vazquez-03,Vazquez-04,Dall'Asta-08}.

A pending task in the study of this class of models for language
competition is, therefore, the detailed analysis of the role of the
prestige and volatility in their general dynamical properties.  In
addition, for the voter model, macroscopic field theory descriptions
\cite{Al_Hammal_Chate-05,Vazquez-08c} as well as macroscopic and
analytical solutions in different complex networks \cite{Vazquez-08a}
have been reported, but there is still a lack of useful macroscopic
descriptions of these models for arbitrary values of the prestige and
volatility parameters.  The general aim of this paper is then, to
explore the behavior of these models for a wide range of these
parameters values, and to derive appropriate macroscopic descriptions
that account for the observed order-disorder (language
dominance-coexistence) transitions in the volatility-prestige
parameter space. In particular, we analyze how the introduction
of an intermediate bilingual state affects language coexistence, by
comparing the regions of coexistence and one-language dominance of the
ASM and BM in the parameter space.  In addition, we study how these
regions are modified, within the same models, when the dynamics takes
place on networks with different topologies.

The paper is organized as follows.  In section \ref{AS-model}, we
introduce and study the Abrams-Strogatz model on fully connected and
complex networks.  Starting from the microscopic dynamics, we derive
ordinary differential equations for the  global magnetization
(difference between the fraction of speakers of each language) and the
interface density (fraction of links connecting opposite-language
speakers).  We use these equations to analyze ordering and stability
properties of the system, that is, whether there is language
coexistence or monolingual dominance in the long time limit.   In
section \ref{MW-model}, we introduce and study the Bilinguals model,
following an approach similar to the one in section \ref{AS-model}.
In section \ref{Square}, we address the behavior of these language
competition models and the order-disorder transitions on square
lattices. In particular, we build a macroscopic description of the
dynamics of the ASM on square lattices by deriving partial
differential equations for the magnetization field, that depend on
space and time.  Finally, in section \ref{Summary} we present a
discussion and a summary of the results.

\section{Abrams and Strogatz model}
\label{AS-model}

The microscopic agent based version \cite{Stauffer-07,Castello-06} of
the model proposed by Abrams and Strogatz \cite{Abrams-03} considers a
population of N individuals sitting in the nodes of a social network
of interactions. Every individual can speak two languages $X$ and $Y$,
but it uses only one at a time.  In a time step, an individual chosen
at random is given the possibility to give up the use of its language
and start using the other language.  The likelihood that the
individual changes language use depends on the fraction of its
neighbors using the opposite language. Neighbors are here understood
as agents sitting in nodes directly connected by a link of the
network.  The language switching probabilities are defined as
\begin{eqnarray}
P(X \to Y) &=& (1-\mathcal S) \, \sigma_y^a ~~~\mbox{and} \nonumber
\\ P(Y \to X) &=& \mathcal S \, \sigma_x^a,
\label{PXY}
\end{eqnarray}
where $\sigma_x (\sigma_y)$ is the density of $X (Y)$ neighboring
speakers of a given individual, $0 \le \mathcal S \le 1$ is the
\emph{prestige} of language $X$, and  $a > 0$ is the \emph{volatility}
parameter.  $\mathcal S$ controls the asymmetry of language change
[$\mathcal S>1/2~(\mathcal S <1/2)$ favoring language $X (Y)$],
whereas $a$ measures the tendency to switch language use.  The case
a=1 is a neutral situation, in which the transition probabilities
depend linearly on the local densities. A high volatility regime
regime exists for $a < 1$, with a probability of changing language
state above the neutral case, and therefore agents change their state
rather frequently. A low volatility regime exists for $a > 1$ with a
probability of changing language state below the neutral case with
agents having a larger resistance or inertia to change their state. 

Having defined the model, we would like to investigate the dynamics
and ultimate fate of  the population, that is, whether all individuals
will agree after many interactions in the use of one language or not.
In order to perform an analytical and numerical study of the evolution
of the system we consider, in an analogy to spin models, $X$ and $Y$
speakers as spin particles in states $s=-1$ (spin down) and $s=1$
(spin up) respectively.  Thus, the state of the system in a given time
can be characterized quite well by two macroscopic quantities:  The
global magnetization $m \equiv \frac{1}{N} \sum_{i=1}^N S_i$, ($S_i$
with $i=1,..,N$ is the state of individual $i$ in a population of size
$N$), and the density of pairs of neighbors using different languages
$\rho \equiv \frac{1}{2 N_l}\sum_{<i j>} (1-S_i S_j)/2$, where $N_l$
is the number of links in the network and the sum is over all pair of
neighbors.  The magnetization measures the balance in the fractions of
$X$ and $Y$
speakers ($m=0$ corresponding to the perfectly balanced case), whereas
$\rho$ measures the degree of disorder in the system.  The case
$|m|=1$ and $\rho = 0$ corresponds to the totally ordered situation,
with all individuals using the same language, while $|m|<1$ and  $\rho
>0$ indicates that the system is disordered, composed by both type of
speakers.

The aim is to obtain differential equations for the time evolution of
the average values of $m$ and $\rho$.  These equations are useful in
the study of the properties of the system, from an analytical point of
view.  We start by deriving these equations in the case of a highly
connected society with no social structure (fully connected network),
that corresponds to the simplified assumption of a ``well mixed''
population, widely used in population dynamics.  We then obtain the
equations in a more realistic scenario, when the topology of
interactions between individuals is a social complex network.  We
shall see that the results depend on the particular properties of the
network under consideration, reflected in the moments of the degree
distribution.

\subsection{Fully connected networks}
\label{Fully}

We consider a network composed by $N$ nodes, in which each node has a
connection to any other node.  In a time step $\delta t = 1/N$, a node
$i$ with state $s$ ($s=\pm 1$) is chosen with  probability
$\sigma_s$. Then, according to the transitions (\ref{PXY}), $i$
switches its sate with probability
\begin{equation}
P(s \to -s) = \frac{1}{2}(1-s v)\left( \sigma_{-s} \right)^a,
\label{Ps-s}
\end{equation}
where $\sigma_{-s}$ is the density of neighbors of $i$ with state
$-s$, that in this fully connected network is equal to the global
density of $-s$ nodes.  Given that the total number of individuals is
conserved we have that $\sigma_- + \sigma_+ =1$.  We define the bias
$v \equiv 1-2\mathcal S$ ($-1<v<1$) as a measure of the preference for
one of the two languages, with $v>0$ ($v<0$) favoring the $s=1$
($s=-1$) state.  In the case that the switch occurs, the density
$\sigma_s$ is reduced by $1/N$, for which the magnetization
$m=\sigma_+ - \sigma_-$ changes by $-2 s/N$.  Then, the average change
in the magnetization can be written as
\begin{equation}
\frac{dm(t)}{dt} = \frac{1}{1/N} \left[ \sigma_- P(- \to +)
  \frac{2}{N} - \sigma_+ P(+ \to -) \frac{2}{N} \right].
\label{dmdt0}
\end{equation}
Using Eq.~(\ref{Ps-s}) and expressing the global densities
$\sigma_{\pm}$ in terms of the magnetization,  $\sigma_{\pm}=(1 \pm
m)/2$, we arrive to
\begin{equation}
\frac{dm(t)}{dt} = 2^{-(a+1)} (1-m^2) \left[ (1+v) (1+m)^{a-1} -(1-v)
  (1-m)^{a-1} \right].
\label{dmdt}
\end{equation}
Equation~(\ref{dmdt}) describes the evolution of a very large system
($N \gg 1$) at the macroscopic level, neglecting finite size
fluctuations.  This equation for the magnetization is enough to
describe the system, given that the density of neighboring nodes in
opposite state $\rho$ can be indirectly obtained through the relation
\begin{equation}
\rho(t) = 2\, \sigma_+(t) \, \sigma_-(t) =
\frac{\left[1-m^2(t)\right]}{2}.
\end{equation}

\subsubsection{Stability}
\label{Fully-stability}

\begin{figure}[t]
\begin{center}
 \includegraphics[width=0.6\textwidth]{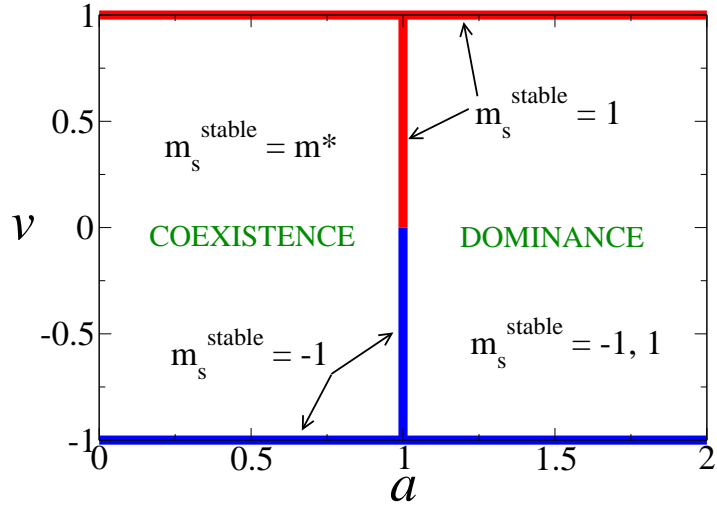}
\end{center}
 \caption{Coexistence and dominance regions of the Abrams-Strogatz
   model in a fully connected network.  For values of the volatility
   parameter $a > 1$, the stable solutions are those of
   language dominance, i.e., all individuals using language $X$
   ($m_s=-1$) or all using language $Y$ ($m_s=1$), whereas for $a<1$
   both languages coexist, with a relative fraction of speakers that
   depends on $a$ and the difference between languages' prestige,
   measured by the bias $v$.  In the extreme case $v=-1$ ($v=1$), only
   language switchings towards $X$ ($Y$) are allowed, and thus only
   one dominance state is stable, independent on $a$.}
 \label{stab-CG}
\end{figure}

Equation~(\ref{dmdt}) has three stationary solutions
\begin{eqnarray}
m_{-}=-1, ~~~ m^*=\frac{(1-v)^{\frac{1}{a-1}}-
  (1+v)^{\frac{1}{a-1}}}{(1-v)^{\frac{1}{a-1}}+(1+v)^{\frac{1}{a-1}}}~~~
\mbox{and}~~~ m_{+}=1.
\label{solutions_AS}
\end{eqnarray}
The stability of each of the solutions depends on the values of the
parameters $a$ and $v$. A simple stability analysis can be done by
considering a small perturbation $\epsilon$ around a stationary
solution $m_s$.  For $m_s=m_{\pm}$, we replace $m$ in Eq.~(\ref{dmdt})
by $m=\pm 1 \mp \epsilon$ (with $\epsilon >0$), and expand to first
order in $\epsilon$ to obtain

\begin{equation}
\frac{d \epsilon}{dt} = 2^{-a} \left[ (1 \mp v) \epsilon^{a-1} -
  2^{a-1} (1 \pm v) \right] \epsilon.
\end{equation}
When $a<1$, $\epsilon^{a-1} \to \infty$ as $\epsilon \to 0$, thus both
solutions $m_{\pm}$ are unstable, whereas for $a>1$, $\epsilon^{a-1}
\to 0$ as $\epsilon \to 0$, thus $m_{\pm}$ are stable.  In the line
$a=1$, $m_{+}$ is unstable (stable) for $v<0$ ($v>0$), and vice-versa
for $m_{-}$.  The same analysis for the intermediate solution $m^*$
leads to
\begin{eqnarray}
\frac{d \epsilon}{dt} = 2^{-(a+1)} (a-1) \left( 1-{m^*}^2 \right)
\left[ (1 + v) (1+m^*)^{a-2} + (1 - v) (1-m^*)^{a-2} \right]
\epsilon. \nonumber \\
\label{}
\end{eqnarray}
Then, $m^*$ is unstable (stable) for $a>1$ ($a<1$). In
Fig.~\ref{stab-CG} we show the regions of stability and instability of
the stationary solutions on the $(a,v)$ plane obtained from the above
analysis. We observe a region of coexistence ($m^*$ stable) and one of
bistable dominance ($m_{+}$ and $m_{-}$ stable).

\begin{figure}[t]
\begin{center}
 \includegraphics[width=0.6\textwidth]{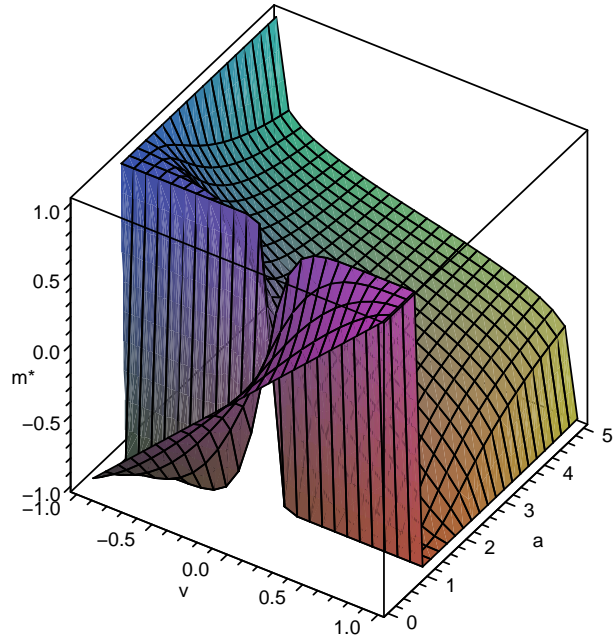}
\end{center}
\caption{ Stationary solution $m^*(a,v)$ for the Abrams-Strogatz model
  (vertical axis) as a function of the two parameters of the model,
  $a$ and $v$ (horizontal-plane). See Expression~(\ref{solutions_AS}).
  Notice how $m^*$ approaches the values of the two trivial stationary
  solutions, $m_{-}=-1$ and $m_{+}=+1$ when $a\rightarrow 1$: for
  $v>0$, $\lim_{a \to 1^\pm} (m^*)=\mp1$. The opposite holds for
  $v<0$. The non-trivial stationary solution, $m^*$, is effectively
  not defined at $a=1$, and in this case the system has only two
  stationary states, $m_{-}$ and $m_{+}$.  The figure illustrates the
  change of stability of $m^*$ at $a_c = 1$.}
 \label{IDL_plot-stab_AS}
\end{figure}

The non-trivial stationary solution, $m^*$, is shown in
Fig.~\ref{IDL_plot-stab_AS} as a function of the parameters $a$ and
$v$. For the coexistence regime ($a<1$), the absolute value of the
stable stationary magnetization $|m^*|$ increases with both, $|v|$ and
$a$. When $v\neq0$ the coexistence solution includes a majority of
agents using the language with higher prestige, and the rest of the
agents using the language with lower prestige.  On the contrary, for
the dominance regime ($a>1$) $|m^*|$ decreases with $a$ and increases
with $|v|$.

\begin{figure}[t]
\begin{center}
 \includegraphics[width=0.6\textwidth]{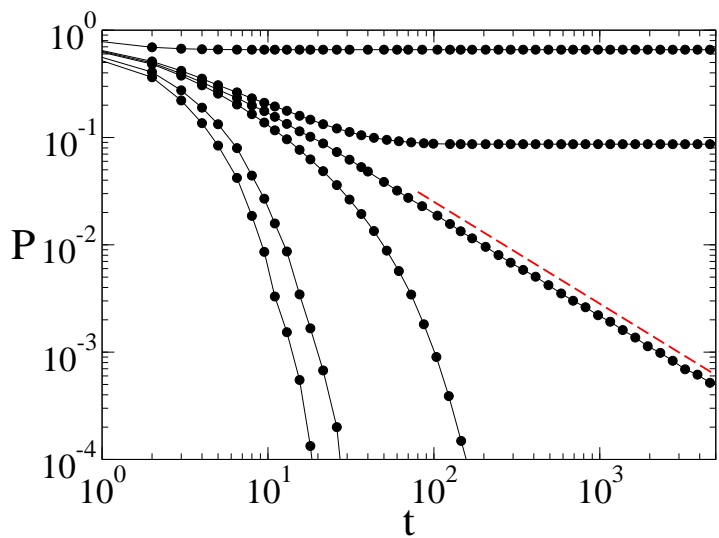}
 \caption{Probability $P(t)$ that the system is still alive at time
   $t$, when it starts from a configuration composed by an up spin in
   a sea of $10^5-1$ down spins, endowed with the Abrams-Strogatz
   dynamics with equivalent languages (bias $v=0$), on a
   fully connected network.  Different curves correspond to the values
   $a=0.90, 0.99, 1.00, 1.01, 1.10$ and $2.0$ (from top to bottom).
   At $a_{c} \simeq 1.0$, $P(t)$ follows a power law decay with exponent
   $\delta \simeq 0.95$, indicated by the dashed line.}
 \label{P-AS}
\end{center}
\end{figure}

In order to account for possible finite size effects neglected in
Eq.~(\ref{dmdt}) we have run numerical simulations in a fully
connected network. We first notice that the solutions $m= \pm 1$
correspond to the totally ordered \emph{absorbing} configurations,
that is, once the system reaches those configurations it never escapes
from them.  This is because, from the transition probabilities
Eq.~(\ref{Ps-s}), a node never flips when it has the same state as all
its neighbors.  Thus, to study the stability of these solutions we
have followed a standard approach \cite{Marro_Dickman_1999} that
consists of adding a defect (seed) to the initial absorbing state and
let the system evolve (spreading experiment).  If, in average, the
defect spreads over the entire system, then the absorbing state is
unstable, otherwise if the defect quickly dies out, the absorbing
state is stable.    For instance, to study the stability of $m=-1$, we
started from a configuration composed by $N-1$ down spins and $1$ up
spin (seed), that corresponds to a magnetization $m=-1+2/N \gtrsim
-1$, and we let the system evolve until an absorbing configuration was
reached.  Whether $m=-1$ is stable or not depends on the values of $v$
and $a$.  If $m=-1$ is unstable, then the seed creates many up spins
and spreads over the system, to end in one of the absorbing states.
If $m=-1$ is stable, then the initial perturbation dies out, and the
system ends in the $m=-1$ absorbing state.  The theory of criticality
predicts that the survival probability $P(t)$, i.e, the fraction of
realizations that have not died up to time $t$, follows a power-law at
the critical point \cite{Marro_Dickman_1999}, where the stability of
the absorbing solution  changes.  Figure~\ref{P-AS} shows that for a
fixed value of the bias $v=0$, $P(t)$ decreases exponentially fast to
zero for values of $a>1$, while it reaches a constant value for $a<1$.
For $a_{c} \simeq 1.0$, $P(t)$ decays as $P(t)\sim t^{-\delta}$, with
$\delta \simeq 0.95$, indicating the transition line from an unstable
to a stable solution $m=-1$ as $a$ is increased, in agreement with the
previous stability analysis.

Following the same procedure, we have also run spreading experiments
to check the stability transition for different values of the bias.
For $v=-0.02$ and $v=-0.2$, on a system of size $N=10^5$, we found the
transitions at $a \simeq 1.007$ and $a \simeq 1.052$, respectively.
These values are slightly different from the analytical value $a_{c}
\simeq 1.0$, but we have verified that as $N$ is increased, the values
become closer to $1.0$, in agreement with the stability analysis on
infinite large systems.

An alternative and more visual way of studying stability in the mean
field limit, is by writing Eq.~(\ref{dmdt}) in the form of a
time-dependent Ginzburg-Landau equation
\begin{equation}
\frac{dm(t)}{dt} = - \frac{\partial V_{a,v}(m)}{\partial m},
\label{TDGL}
\end{equation}
with potential
\begin{eqnarray}
V_{a,v}(m) &\equiv& 2^{-a} \Biggl\{-v m - \frac{1}{2}(a-1) m^2 +
\frac{v}{6} \left[ 2-(a-1)(a-2) \right] m^3 \nonumber \\ &+&
\frac{1}{24} (a-1) \left[6-(a-2)(a-3) \right] m^4 +\frac{v}{10}
(a-1)(a-2)m^5 \nonumber \\ &+& \frac{1}{36}(a-1)(a-2)(a-3) m^6
\Biggr\}.
\label{Vav}
\end{eqnarray}
$V_{a,v}$ is obtained by Taylor expanding the term in square brackets
of Eq.~(\ref{dmdt}) up to $3$-rd order in $m$, and integrating once
over $m$.  We assume that higher order terms in the expansion are
irrelevant, and the dynamics is well described by an $m^6$-potential.

Within this framework, the state of the system, represented by a point
$m(t)$ in the magnetization one-dimensional space $-1<m<1$, moves
``down the potential hill'', trying to reach a local minimum.
Therefore, a minimum of  $V_{a,v}$ at some point $m_s$ represents a
stable stationary solution, given that if the system is moved apart
from $m_s$ and then released, it immediately goes back to $m_s$,
whereas a maximum of $V_{a,v}$ represents an unstable stationary
solution.
\begin{figure}[t]
\begin{center}
 \includegraphics[width=0.60\textwidth]{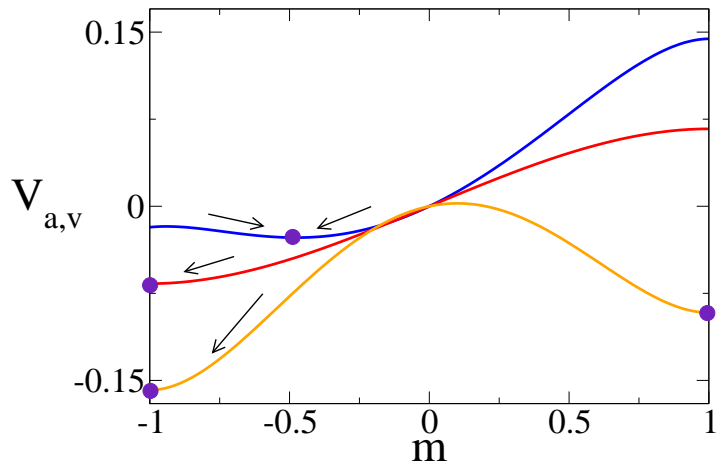}
 \caption{Ginzburg-Landau potential from Eq.~(\ref{Vav}), for the
   Abrams-Strogatz model with bias $v=-0.1$ and values of volatility
   $a=0.8,1.0$ and $2.0$ (from top to bottom).  Arrows show the
   direction of the system's magnetization towards the stationary
   solution (solid circles).  For $a=0.8$ the minimum is around $m
   \simeq -0.5$, indicating that the system relaxes towards a
   partially ordered stationary state, while for $a=1.0$ and $2.0$, it
   reaches the complete ordered state $m=-1$.}
 \label{V-av}
\end{center}
\end{figure}
As Fig.~\ref{V-av} shows, for $a<1$ and all values of $v$, the
single-well potential has a minimum at $|m_s|<1$ ($m_s\simeq -0.5$ for
$a=0.8$ and $v=-0.1$), indicating that the system reaches a partially
ordered stable state, with fractions $0.75$ and $0.25$ of down and up
spins, respectively, and a density of opposite-state links $\rho
\simeq 0.375$. For $a>1$, the double-well potential has a minimum at
$m=\pm 1$, thus depending on the initial magnetization, the system is
driven to one of the stationary solutions $m=\pm 1$, corresponding to
the totally ordered configurations in which $\rho=0$.

This description works well in infinite large systems, where finite
size fluctuations are zero.  But in finite systems, the absorbing
solutions $m=\pm 1$ are the only ``truly stationary states'', given
that fluctuations ultimate take the system to one of those states.
Even for the case $a<1$, where the minimum is at $|m_s|<1$, the
magnetization fluctuates around $m_s$ for a very long time until after
a large fluctuation it reaches $|m|=1$, and the system freezes.

\subsubsection{$a=1$ case: the voter model}
\label{Fully-voter}

\begin{figure}[t]
\begin{center}
 \includegraphics[width=0.60\textwidth]{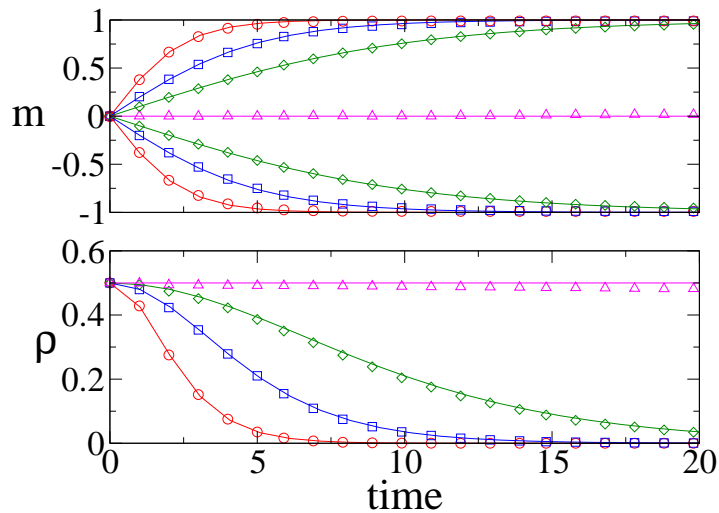}
 \caption{Abrams-Strogatz model on a fully connected network of
   $N=1000$ nodes with volatility $a=1$.  Upper panel: Average
   magnetization $m$ vs time for values of the bias $v=0.8, 0.4, 0.2,
   0.0, -0.2, -0.4$ and $-0.8$ (from top to bottom).  Lower panel:
   Average density of opposite-state links $\rho$ vs time for $v=0.0,
   0.2, 0.4$ and $0.8$ (top to bottom).  Open symbols are the results
   from numerical simulations, while solid lines in the upper and
   lower panels correspond to the solutions from Eqs.~(\ref{m}) and
   (\ref{r}) respectively.  Averages are over $100$ independent
   realizations starting from a configuration with a uniform
   distribution of spins and global magnetization $m(0)=0$.}
 \label{m-r-t}
\end{center}
\end{figure}

For $a=1$ the ASM becomes equivalent to the voter model.  A switching
probability proportional to the local density of neighbors in the
opposite state is statistically equivalent to adopt the state of a
randomly chosen neighbor.  In this limit of neutral volatility, $a=1$,
Eq.~(\ref{dmdt}) becomes
\begin{equation}
\frac{dm}{dt} = \frac{v}{2} (1-m^2),
\end{equation}
whose solution is
\begin{equation}
m(t) = \frac{(1+m_0) e^{v t}-(1-m_0)}{(1+m_0) e^{v t}+(1-m_0)},
\end{equation}
with $m_0=m(t=0)$.  For a uniform initial condition is $m_0=0$, thus
\begin{equation}
m(t) = \tanh(v t /2),
\label{m}
\end{equation}
and
\begin{equation}
\rho(t) = \frac{1}{2} \left[ 1-\tanh^2(v t /2) \right].
\label{r}
\end{equation}
In Fig.~\ref{m-r-t} we observe that the analytical solutions from
Eqs.~(\ref{m}) and (\ref{r}) agree very well with the results from
numerical simulations of the model, for large enough systems, and they
also reproduce the Monte Carlo results found in \cite{Stauffer-07}.
This is so, because finite-size fluctuations effects are negligible
compare to bias effects, even for a small bias.

When the bias is exactly zero, one obtains that in an infinite large
network $dm/dt = 0$, thus $m$ and $\rho$ are conserved.  However, in a
finite network, fluctuations always lead the system to one of the
absorbing states \cite{Stauffer-07}.  To find how the  system relaxes
to the final state, one needs to calculate the evolution of the second
moment $\langle m^2 \rangle$ of the magnetization, related to the
fluctuations in $m$, where the symbol $\langle ~ \rangle$ represents
an average over many realizations.  This leads to a decay of the
average density of opposite-state links of the form (see
\cite{Vazquez-08a})
\begin{equation}
\langle \rho(t) \rangle = \frac{1}{2}\left[1-\langle m^2(t) \rangle
  \right] = \langle \rho(0) \rangle~ e^{-2 t/N}.
\end{equation}

In terms of the potential description of Eq.~(\ref{TDGL}), we observe
that when $v \neq 0$, $V_{a,v}$ has only one minimum (see
Fig.~\ref{V-av}), thus the system has a preference for one of the
absorbing states only, whereas if $v =0$, is $V_{a,v}=0$, and the
magnetization is conserved ($m(t)=m(0)=$ constant).  In finite
systems, even though the average magnetization over many realizations
is conserved, the system stills orders in individual realizations by
finite size fluctuations.

\subsection{Complex networks}
\label{AS-nets}

In real life, most individuals in a large society interact only with a
small number of acquaintances, and they all form a social network of
connections, where nodes represent individuals and links between them
represent interactions.  Thereby, we consider a network of $N$ nodes,
with a given degree distribution $P_k$, representing the fraction of
individuals connected to $k$ neighbors, such that $\sum_k P_k=1$.  In
order to develop a  mathematical approach that is analytically
tractable, we assume that the network has no degree correlations, as
it happens for instance in  Erd\"os-Renyi \cite{ER} and
Barab\'asi-Albert scale-free networks \cite{Barabasi_1999}.  It turns out 
that dynamical correlations between the states of second nearest-neighbors 
are negligible in voter models on uncorrelated networks
\cite{Vazquez-08a,Castellano-09-b}.  Thus, taking into account only
correlations between first nearest-neighbors allows to use an approach, 
called \emph{pair approximation}, that leads to analytical results in good 
agreement with simulations.  In this section, we shall use this 
approximation to build equations for the magnetization and the density of
links in opposite state.

In a time step $\delta t = 1/N$, a node $i$ with degree $k$ and state $s$
is chosen  with  probability $P_k \, \sigma_s$. Here we assume that
the density of nodes in state $s$ within the subgroup of nodes with
degree $k$ is independent on $k$ and equal to the  global density
$\sigma_s$.  Then, according to transitions (\ref{PXY}), $i$ switches
its sate with probability
\begin{equation}
P(s \to -s) = \frac{1}{2}(1-s v)\left( n_{-s}/k \right)^a,
\label{Ps-s1}
\end{equation}
where we denote by $n_{-s}$ the number of neighbors of $i$ in the
opposite state $-s$ ($0 \le n_{-s} \le k$) and the bias $v$ is defined
as in the previous section.  If the switch occurs, the density
$\sigma_s$ is reduced by $1/N$, for which the magnetization
$m=\sigma_+ - \sigma_-$ changes by $-2 s/N$, while the density $\rho$
changes by $2(k-2 n_{-s})/\mu N$, where $\mu \equiv \sum_k k P_k$ is
the average degree of the network.
Thus, in analogy to section \ref{Fully}, but now plugging the
transition probabilities from Eq.~(\ref{Ps-s1}) into
Eq.~(\ref{dmdt0}), we write the average change in the magnetization as
\begin{eqnarray}
\frac{d m(t)}{d t} &=& \sum_k \frac{P_k  \sigma_-}{1/N}
\sum_{n_{+}=0}^k B(n_{+},k) \frac{(1+v)}{2} \left( \frac{n_{+}}{k}
\right)^a \frac{2}{N} \nonumber \\ &-& \sum_k \frac{P_k \sigma_+}{1/N}
\sum_{n_{-}=0}^k B(n_{-},k) \frac{(1-v)}{2} \left( \frac{n_{-}}{k}
\right)^a \frac{2}{N},
\label{dmdt1}
\end{eqnarray}
and similarly, the change in the density of links in opposite state as
\begin{eqnarray}
\frac{d \rho(t)}{dt} &=& \sum_k \frac{P_k \sigma_-}{1/N}
\sum_{n_{+}=0}^k B(n_{+},k) \frac{(1+v)}{2} \left( \frac{n_{+}}{k}
\right)^a~ \frac{2 (k-2 n_{+})}{\mu N} \nonumber \\ &+& \sum_k
\frac{P_k  \sigma_+}{1/N} \sum_{n_{-}=0}^k B(n_{-},k) \frac{(1-v)}{2}
\left( \frac{n_{-}}{k} \right)^a~ \frac{2 (k-2 n_{-})}{\mu N}.
\label{drdt1}
\end{eqnarray}
We denote by $B(n_{s},k)$, the probability that a node of degree $k$
and state $-s$ has $n_{s}$ neighbors in the opposite state $s$.
Defining the $a$-th moment  of $B(n_{s},k)$ as
\begin{eqnarray*}
\langle n_{s}^a \rangle_k \equiv \sum_{n_{s}=0}^k B(n_{s},k) n_{s}^a,
\end{eqnarray*}
we arrive to the equations
\begin{equation}
\frac{d m(t)}{d t} = \sum_k \frac{P_k} {2\,k^a} \left[ (1+v) (1-m)
  \langle n_+^a \rangle_k - (1-v) (1+m) \langle n_-^a \rangle_k
  \right],
\label{dmdt2}
\end{equation}
\begin{eqnarray}
\frac{d \rho(t)}{dt} &=& \sum_k \frac{P_k}{2 \, \mu \, k^{a}} \Big\{
(1+v) (1-m) \left[ k \langle n_+^a \rangle_k - 2 \langle n_+^{(1+a)}
  \rangle_k \right] \nonumber \\  &+& (1-v) (1+m) \left[ k \langle
  n_-^a \rangle_k -  2 \langle n_-^{(1+a)} \rangle_k \right] \Big\}.
\label{drdt2}
\end{eqnarray}

\subsubsection{$a=1$ case: the voter model}

In order to develop an intuition about the temporal behavior of $m$
and $\rho$ from Eqs.~(\ref{dmdt2}) and (\ref{drdt2}), we first analyze
the simplest and non-trivial case $a=1$, that corresponds to the voter
model on complex networks.  A rather complete analysis of the time
evolution and consensus times of this model on uncorrelated networks,
for the symmetric case $v=0$, can be found in
\cite{Vazquez-08a}. Following a similar approach, here we study the
general situation in which the bias $v$ takes any value.  To obtain
closed expressions for $m$ and $\rho$, we consider that the system is
``well mixed'', in the sense that the different types  of links are
uniformly distributed over the network.  Therefore, we assume that the
probability that a link picked at random is of  type $+ -$ is equal to
the global density of $+ -$ links $\rho$. Then, $B(n_{-s},k)$ becomes
the binomial distribution with
\begin{equation}
P(-s|s)=\rho/2 \sigma_s
\label{P-ss}
\end{equation}
as the single event probability that a first nearest-neighbor of a
node with state $s$ has state $-s$.  Here, we use the fact that
in uncorrelated networks dynamical correlations between the states of
second nearest-neighbors vanish (pair approximation).  $P(-s|s)$ is
calculated as the ratio between the total number of links $\rho \mu
N/2$ from nodes in state $s$ to nodes in state $-s$, and the total
number of links $N \sigma_s \mu$ coming out from nodes in state $s$.
Taking $a=1$ in Eqs.~(\ref{dmdt2}) and (\ref{drdt2}), and replacing
the first and second moments of $B(n_{-s},k)$ by
\begin{eqnarray*}
\langle n_{-s} \rangle &=& P(-s|s) k, \\ \langle n_{-s}^2 \rangle &=&
P(-s|s) k + P(-s|s)^2 k(k-1),
\end{eqnarray*}
leads to the following two coupled closed equations for $m$ and $\rho$
\begin{eqnarray}
  \label{dmdt3}
  \frac{d m(t)}{d t}&=&v \rho \\
  \label{drdt3}
  \frac{d \rho(t)}{d t}&=&\frac{\rho}{\mu} \Bigg\{ \mu-2 - \frac{2
    (\mu-1) (1+v\,m) \rho}{(1-m^2)} \Bigg\}.
\end{eqnarray}
For $v=0$, the above expressions agree with the ones of the symmetric
voter model \cite{Vazquez-08a}.  For the asymmetric case $v \ne 0$, we
have checked numerically that the only stationary solutions are $(m=1,
\rho=0)$ for $v>0$ and $(m=-1, \rho=0)$ for $v<0$, that correspond to
the fully  ordered state, as we were expecting.  Even though an exact
analytical solution of Eqs.~(\ref{dmdt3}) and (\ref{drdt3}) is hard to
obtain, we can still find a solution in the long time limit, assuming
that $\rho$ decays to zero as
\begin{equation}  \rho = A\, e^{-t/2\tau(v)},~~~\mbox{for}~~t \gg 1,
\label{r6}
\end{equation} 
where $A$ is a constant given by the initial state and $\tau(v)$ is
another constant that depends on $v$, and quantifies the rate of decay
towards the solutions $m=1$ or $m=-1$.  To calculate the  value of
$\tau$ we first replace the ansatz from Eq.~(\ref{r6}) into
Eq.~(\ref{dmdt3}), and solve for $m$ with the boundary conditions
$m(\rho=0) = 1$ and $-1$, for $v>0$ and $v<0$, respectively.  We
obtain
\begin{eqnarray}
m = \left\{ \begin{array}{ll} 1-2 v \tau \rho& \mbox{if $v >
    0$};\\ -1- 2 v \tau \rho & \mbox{if $v < 0$}.\end{array} \right.  
\end{eqnarray}
Then, to first order in $\rho$ is
\begin{eqnarray}
(1-m^2) = \left\{ \begin{array}{ll} 4 v \tau \rho& \mbox{if $v >
      0$};\\ - 4 v \tau \rho & \mbox{if $v < 0$}.\end{array} \right.
\label{1m2}
\end{eqnarray}
 Replacing the above expressions for $m$ and $(1-m^2)$ into Eq.~(\ref{drdt3}),
 and keeping only the leading order terms, we arrive to the following
 expression for $\tau$
\begin{eqnarray}
\tau(v) = \left\{ \begin{array}{ll} \frac{\mu-1-v}{2 v (\mu-2)} &
  \mbox{if $v > 0$};\\ \frac{1- \mu - v}{2 v (\mu-2)} & \mbox{if $v <
    0$}.\end{array} \right.
\label{tau}
\end{eqnarray}
Finally, the magnetization for long times behave as
\begin{eqnarray}
m = \left\{ \begin{array}{ll} 1-\frac{(\mu - 1-v) A}{\mu-2}
  \exp{\left[-\frac{v (\mu-2)}{\mu - 1 - v}t\right]}& \mbox{if $v >
    0$};\\ -1 + \frac{(\mu- 1 + v) A}{\mu-2} \exp{\left[\frac{v
        (\mu-2)}{\mu- 1 + v}t\right]} & \mbox{if $v < 0$}, \end{array}
\right.
\label{m6}
\end{eqnarray}
whereas the density of opposite state links decays as
\begin{equation}  \rho = A\, \exp{\left[-\frac{|v| (\mu-2)}{\mu- 1 -
        |v|}t \right]}.
\label{r7}
\end{equation} 
Using the expression for $\tau(S)$ from Eq.~(\ref{tau}) in
Eq.~(\ref{1m2}), and taking the limit $v \to 0$, we find that $\rho(t)
= \frac{(\mu-2)}{2(\mu-1)}\left[ 1-m(t)^2 \right]$, in agreement with
previous results of the voter model on uncorrelated networks
\cite{Vazquez-08a}.  By taking $\mu = N-1 \gg 1$ in Eqs.~(\ref{m6})
and (\ref{r7}), we recover the expressions for $m$ and $\rho$ on fully
connected networks [Eqs.~(\ref{m}) and (\ref{r}), respectively], in
the long time limit.  This result means that the evolution of $m$ and
$\rho$ in the biased voter model on uncorrelated networks is very
similar to the mean-field case, with the time rescaled by the constant
$\tau$ that depends on the topology of the network, expressed by the
mean connectivity $\mu$.  From the above equations we observe that the
system reaches the dominance state $\rho=0$ in a time of order $\tau$.
For the special case $v=0$, $\tau$ diverges, thus Eqs.~(\ref{m6}) and
(\ref{r7}) predict that both $m$ and $\rho$ stay constant over time.
However, as mentioned in section \ref{Fully-voter}, finite-size
fluctuations drive the system to the absorbing state ($\rho=0,
|m|=1$).  Taking fluctuations into account, one finds that the
approach to the final state is described by the decay of the average
density $\rho$ \cite{Vazquez-08a} 
\begin{equation}
\langle \rho(t) \rangle = \frac{(\mu-2)}{2(\mu-1)} e^{-2 t/T},
\end{equation}
where $T \equiv \frac{(\mu-1) \mu^2 N}{(\mu-1) \mu_2}$, depends on the
system size $N$, and the first and second moments, $\mu$ and $\mu_2$
respectively, of the network.

\subsubsection{Stability analysis}
\label{diagram_AS-in_RDR}

\begin{figure}[t]
\begin{center}
 \includegraphics[width=0.60\textwidth]{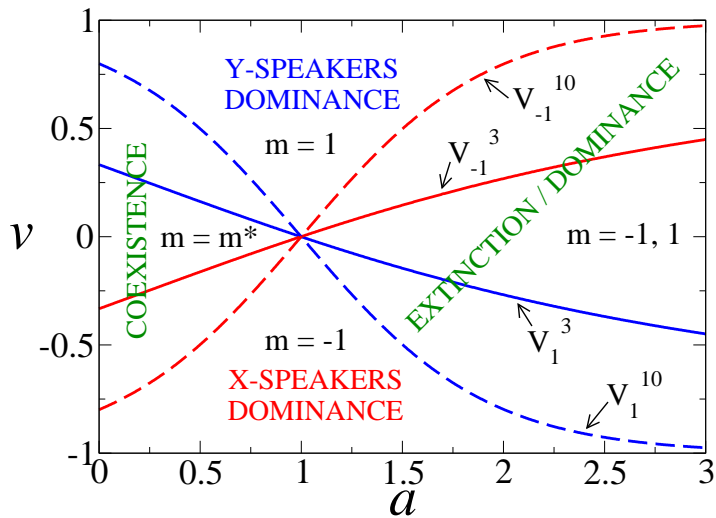}
 \caption{Stability diagram for the Abrams-Strogatz model on a
   degree-regular random graph, obtained by numerical integration of
   Eqs.~(\ref{dmdt2}), (\ref{drdt2}) and (\ref{moments}).  The
   solution $m=1$ is stable above the line $\mathcal V_1$, while the
   solution $m=-1$ is stable below the line $\mathcal V_{-1}$.  Solid
   and dashed lines correspond to graphs with degrees $\mu=3$ and
   $\mu=10$ respectively.  In the coexistence region, where the stable
   solution is $m^*$, the system is composed by both type of users,
   while in the dominance region, users of either one or the other
   language prevail, depending on the initial state.  We observe that
   the region of coexistence is reduced, compared to the model on
   fully connected networks (Fig.~\ref{stab-CG}), and that there are
   also two single-dominance regions where always the same language
   dominates.}
 \label{stab-DR}
\end{center}
\end{figure}

As in fully connected networks, we assume that Eq.~(\ref{dmdt2}) for
the magnetization has three stationary solutions.  Indeed, we have
numerically verified that for different types of networks there is,
apart from the trivial solutions $m=1,-1$, an extra non-trivial
stationary solution $m=m^*$.  Due to the rather complicated form of
Eq.~(\ref{dmdt2}), we try to study the stability of the solutions in
an approximate way, and find a qualitative picture of the stability
diagram in the $(a,v)$ plane.   For the general case in which $a$ and
$v$ take any values, we assume, as in the voter model case, that
$B(n_{-s},k)$ is a binomial probability distribution with single event
probabilities given by Eq.~(\ref{P-ss}).  Then, the explicit form for
the $a$-th moment of $B(n,k)$ is
\begin{equation}
\langle n_{s}^a \rangle = \sum_{n_{s}=0}^k n^a C_{n_{s}}^k
\left(\frac{\rho}{2 \sigma_{-s}} \right)^{n_{s}} \left(
1-\frac{\rho}{2 \sigma_{-s}} \right)^{k-n_{s}}.
\label{moments}
\end{equation}
We also assume that, as it happens for the voter model case $a=1$
[see Eq.~(\ref{1m2})], $\rho$ and $m$ are related by  $\rho(t) \simeq
\frac{q}{2} \left[1-m^2(t)\right]$, where $q$ is a constant that
depends on $a$ and $v$.  We note that this relation satisfies the
fully-ordered-state condition $\rho=0$ when $|m|=1$.  We shall see
that the exact functional form of $q=q(a,v)$ is irrelevant for the
stability results, as long as $q>0$.  To simplify calculations even
more, we consider that the network is a degree-regular random graph
with degree distribution $P_k = \delta_{k,\mu}$, that is, all nodes
have exactly $\mu$ neighbors chosen at random.  Then, replacing the
above expression for the moments into Eq.~(\ref{dmdt2}), and
substituting $\rho$ by the approximate value $\frac{q}{2} \left[1-m^2
  \right]$, we arrive to the following closed equation for $m$
\begin{eqnarray}
\frac{dm}{dt} &=& \frac{(1-m^2)}{2 \mu^a} \sum_{n=0}^\mu C_{n}^\mu\,
n^a\, \left( \frac{q}{2} \right)^n \Big\{ (1+v) (1+m)^{n-1}
\left[1-p(1+m)\right]^{\mu-n} \nonumber \\ &-&(1-v) (1-m)^{n-1}
\left[1-p (1-m) \right]^{\mu-n} \Big\},
\label{dmdt4}
\end{eqnarray}
where mute indices $n_-$ and $n_+$ were replaced by the index $n$.  To
check the stability of $m=1$, we take $m=1-\epsilon$ in
Eq.~(\ref{dmdt4}), and expand it to first order in $\epsilon$.  We
obtain after some algebra
\begin{equation}
\frac{d \epsilon}{dt} = \frac{\mu^{-a}}{2} \left( \langle n
\rangle_{q} + \langle n^a \rangle_{q} \right) \left[ \mathcal V_1(a) -
  v \right] \epsilon,
\end{equation}
where the symbols $\langle ~ \rangle_{q}$ represent the moments of a
Binomial distribution with probability $q$, and the bias function
$\mathcal V_1(a)$ is defined as
\begin{equation}
\mathcal V_1(a) = \frac{\langle n \rangle_q -  \langle n^a \rangle_q}
         {\langle n \rangle_q + \langle n^a \rangle_q}.
\end{equation}
Then, for a fixed value of $a$ the solution $m=1$ is stable
(unstable), when $v$ is larger (smaller) than $\mathcal V_1(a)$.  The
shape of the function $\mathcal V_1(a)$ can be guessed using that for
$a$ larger (smaller) than $1$, the moment $\langle n^a \rangle$ is
larger (smaller) than $\langle n \rangle$.  Then $\mathcal V_1(a)$
goes to $(\langle n \rangle -1)/(\langle n \rangle +1) \lesssim 1$ and
$-1$ as $a$ approaches to $0$ and $\infty$, respectively.  Also
$\mathcal V_1(a)=0$, for $a=1$.       With a similar stability
analysis we obtained that $m=-1$ is stable (unstable) for the points
$(a,v)$ below (above) the transition line  $\mathcal V_{-1}(a)=-
\mathcal V_1(a)$, while $m=m^*$ is stable in the region where both
$m=-1$ and $m=1$ are unstable.  In Fig.~\ref{stab-DR} we show a
picture that summarizes the stability regions defined by the
transition lines $\mathcal V_1(a)$ and $\mathcal V_{-1}(a)$.  These
lines were obtained by integrating numerically the two coupled
Eqs.~(\ref{dmdt2}) and (\ref{drdt2}), with the moments defined in
Eq.~(\ref{moments}), and finding the points $(a,v)$ where the
stationary solutions $m=1,-1$ became unstable.  We considered two
degree-regular random graphs with degrees $\mu=3$ [solid lines
  $\mathcal V_1^3(a)$ and $\mathcal V_{-1}^3(a)$] and $\mu=10$
[dashed-lines $\mathcal V_1^{10}(a)$ and $\mathcal V_{-1}^{10}(a)$],
thus we took $P_k = \delta_{k,\mu}$ in the equations.  For clarity,
only the stable solutions are labeled in the picture.  We observe that
as the degree of the network increases, the coexistence region expands
and approaches to the corresponding region $a<1$ on fully connected
networks.

\begin{figure}
\begin{center}
\includegraphics[width=0.6\textwidth]{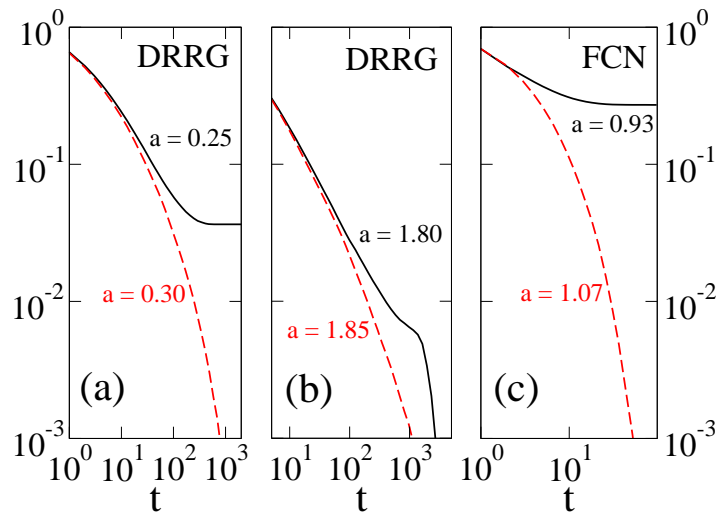}
\caption{Spreading experiments: probability $P(t)$ that the system is
  still alive at time $t$ in the Abrams-Strogatz model with bias
  $v=-0,2$ and various values of volatility $a$, showing the stability
  of the solutions $m=1,-1$.  Dashed curves decay quickly to zero,
  indicating that the solution is stable while solid curves represent
  unstable solutions. (a) Degree-regular random graph
  (DRRG). Stability of the solution $m=-1$: $a=0.25$ (solid curve),
  $a=0.30$ (dashed curve). (b) Degree-regular random graph
  (DRRG). Stability of the solution $m=+1$: $a=1.80$ (solid curve),
  $a=1.85$ (dashed curve).  (c) Fully connected network (FCN).
  Stability of the solutions $m=\pm 1$: $a=0.93$ (solid curve),
  $a=1.07$ (dashed curve).  All curves correspond to an average over
  $10^5$ independent realizations on networks with $N=10^5$ nodes.}
\label{P-AS-FCN-DRRG}
\end{center}
\end{figure}

In order to give numerical evidence, from Monte Carlo simulations, of
the  different phases and transition lines predicted in
Fig.~\ref{stab-DR}, we have run spreading experiments as explained in
section \ref{Fully}, for a degree-regular random graph (DRRG) with
degree $\mu=3$ and $N=10^5$ nodes, and tested the stability of the
homogeneous solutions $m=\pm 1$.  
We
first set the bias in $v=0$ and, by varying $a$, we obtained a
transition at $a_{c} \simeq 1.0$ from dominance to coexistence, as $a$
is decreased:  in the dominance region the survival
probability $P(t)$ decays exponentially fast to zero, indicating that
$m=1$ is stable, while in the coexistence region $P(t)$ reaches a
constant value larger than zero, showing that $m=1$ is unstable (not
shown).  
This transition is the same as the one in
fully connected networks (FCN) (Fig.~\ref{P-AS}).  We then repeated
the experiment with $v=-0.2$, whose results are summarized in
Fig.~\ref{P-AS-FCN-DRRG}, where we show $P(t)$ for different values of
$a$.  Increasing $a$ from $0$, which  corresponds to the coexistence
regime ($m=\pm1$ are unstable solutions, and $m^*$ is stable), we show
in Fig.~\ref{P-AS-FCN-DRRG}(a) how in a DRRG $m=-1$ changes from
unstable to stable at a value $0.25 < a < 0.3$, as $P(t)$ starts to
decay to zero.  This corresponds to crossing the line  $\mathcal
V_{-1}^{3}$ in the horizontal direction (see Fig.~\ref{stab-DR}), and
entering the monostable region where there exist only two solutions,
$m=-1$ stable, and $m=+1$ unstable ($m^*$ becomes equal to $-1$ along
the transition line $\mathcal V_{-1}^{3}$).  In
Fig.~\ref{P-AS-FCN-DRRG}(b) we observe how in a DRRG $m=+1$ becomes
stable at a value $1.80 < a < 1.85$.  This corresponds to crossing the
line $\mathcal V_{1}^{3}$ (see Fig.~\ref{stab-DR}) and entering to the
dominance region, where both $m=\pm 1$ are stable.  Notice that for
$a=1.80$, $P(t)$ first curves up and then it quickly decays to zero at
a time $t \simeq 4000$. This means that a finite fraction of
realizations starting from a system with a single down spin took, in
average, a mean time $t \simeq 4000$ to end up in a  configuration
with all down spins, showing that $m=-1$ is a stable solution.  This
supports our claim that in the monostable region there exist only two
solutions, $m=-1$ stable, and $m=+1$ unstable.  These results confirm
the existence of a quite broad single-dominance region in DRRG ($0.30
\lesssim a \lesssim 1.85$ for $v=-0.2$ and $\mu=3$), in agreement with
the stability diagram obtained in  Fig.~(\ref{stab-DR}), while this
region seems to be absent in FCN.  Indeed, Fig.~\ref{P-AS-FCN-DRRG}(c)
shows how this unstable-stable transition happens in a FCN at a value
$0.93 < a < 1.07$, in agreement with the transition line $a_{c} \simeq
1.0$ in FCN.  Here, both $m=\pm 1$ gain stability at the same point,
and the system enters to the dominance region (see Fig~\ref{stab-CG}).  

In summary, we find that, compared to the fully connected case, the
region of coexistence is shrunk for $v\neq0$, as there appear two
regions where only one solution is stable. These regions also reduce
part of the dominance region.  The effect of the bias is shown to be
more important in DRRGs with low connectivity $\mu$ and, as a general
result, coexistence becomes harder to achieve in sparse networks.

\section{Bilinguals Model}
\label{MW-model}

This model can be regarded as an extension of the Abrams-Strogatz
model in which, besides monolingual users $X$ and $Y$, there is a
third class of individuals that use both languages, that is, bilingual
users labeled with state $Z$.  A monolingual $X$ ($Y$) becomes a
bilingual with a rate depending on the number of its neighbors that
are monolinguals $Y$ ($X$), while direct transitions from one class of
monolingual to the other are forbidden.  This reflects the fact that
individuals that use one language only, are forced to start using both
languages if they want to have a conversation with monolingual users
of the opposite language.  For a similar reason, the transition from a
bilingual $Z$ to a monolingual $X$ ($Y$) depends on the number
of neighbors using language $X$ ($Y$),  which includes bilingual
agents.  Thus, the transition
probabilities between states are given by 
\begin{eqnarray}
P(X \to Z) &=& (1-\mathcal S) \, \sigma_y^a, \nonumber \\ P(Z \to Y)
&=& (1- \mathcal S) \, (1-\sigma_x)^a, \nonumber \\ P(Y \to Z) &=&
\mathcal S \, \sigma_x^a, \nonumber \\ P(Z \to X) &=& \mathcal S \,
(1-\sigma_y)^a,
\label{PXYZ}
\end{eqnarray}
where $\sigma_x$, $\sigma_y$ and $\sigma_z$ are the densities of
neighboring speakers in states $X$, $Y$ and $Z$ respectively, and
$\mathcal S$ is the prestige of language $X$.

As in the ASM, it is convenient to consider monolinguals $X$ and $Y$,
as particles with opposite spins $-1$ and $1$ respectively.
Bilinguals are considered as spin-$0$ particles because they are a 
combination
of the two opposite states.  Given that the model is invariant under the
interchange of $-1$ and $1$ particles, the system is better described
using the global magnetization $m \equiv \sigma_+ - \sigma_-$ and the
density of bilinguals $\sigma_0$, where $\sigma_-$,$\sigma_0$,
$\sigma_+$, are the global densities of nodes in states $-1$, $0$ and
$1$, respectively.  Another alternative could be the use of the
density of connections between different states $\rho \equiv 2
\sigma_- \sigma_+ + 2 \sigma_- \sigma_0 + 2 \sigma_+ \sigma_0$, but
numerical simulations show that $\rho$ and $\sigma_0$ are
proportional.  We now study the evolution of the system on fully
connected and complex networks, by writing equations for $m$ and
$\sigma_0$.

\subsection{Fully connected networks}

In the fully connected case, the local densities of neighbors in the
different states agree with the global densities
$\sigma_-$,$\sigma_0$, $\sigma_+$, thus, using the transition
probabilities Eqs.~(\ref{PXYZ}), the rate equations for $\sigma_-$ and
$\sigma_+$ can be written as
\begin{eqnarray}
\frac{d \sigma_-}{dt} &=& \frac{(1-v)}{2} \sigma_0 (1-\sigma_+)^a -
\frac{(1+v)}{2} \sigma_- \sigma_+^a, \\ \frac{d \sigma_+}{dt} &=&
\frac{(1+v)}{2} \sigma_0 (1-\sigma_-)^a - \frac{(1-v)}{2} \sigma_+
\sigma_-^a,
\end{eqnarray}
where $v \equiv 1-2 \mathcal S$ is the  bias.  The rate equations for
$m=\sigma_+-\sigma_-$ and $\sigma_0=1-\sigma_+-\sigma_-$ can be
derived from the above two equations, and by making the substitutions
$\sigma_s = (1-\sigma_0 + s\, m)/2$, with $s=\pm 1$.  We obtain
\begin{eqnarray}
\label{dmdt5}
\frac{dm}{dt} &=& 2^{-(2+a)} \Big\{ 2 \sigma_0 \left[ (1+v)
  (1+\sigma_0+m)^a - (1-v) (1+\sigma_0 -m)^a \right] \\ &+& (1+v)
(1-\sigma_0-m) (1-\sigma_0 +m)^a - (1-v) (1-\sigma_0+m)
(1-\sigma_0-m)^a \Big\} \nonumber 
\end{eqnarray}
and
\begin{eqnarray}
\label{dsdt}
\frac{d\sigma_0}{dt} &=& 2^{-(2+a)} \Big\{ -2 \sigma_0 \left[ (1+v)
  (1+\sigma_0+m)^a +(1-v) (1+\sigma_0 -m)^a \right] \\ &+& (1+v)
(1-\sigma_0-m) (1-\sigma_0 +m)^a + (1-v) (1-\sigma_0+m)
(1-\sigma_0-m)^a \Big\}. \nonumber
\end{eqnarray}

Equations~(\ref{dmdt5}) and (\ref{dsdt}) are difficult to integrate
analytically, but an insight on its qualitatively behavior can be
obtained by studying the stability of the stationary solutions with
$a$ and $v$.  As in the ASM, we expect that, for a given $v$, an
order-disorder transition appears at some value $a_c$ of the
volatility parameter, where the stability of the stationary solutions
changes.  If $a$ is small, then flipping rates are high, thus we expect
the system to remain in an active disordered state, while for large
enough values of $a$ spins tend to be aligned, thus the system should
ultimately reach full order.  We now calculate the
transition point for the symmetric case $v=0$, and then find an approximate
solution for the linear case $a=1$.

\begin{figure}[t]
\begin{center}
 \includegraphics[width=0.5\textwidth]{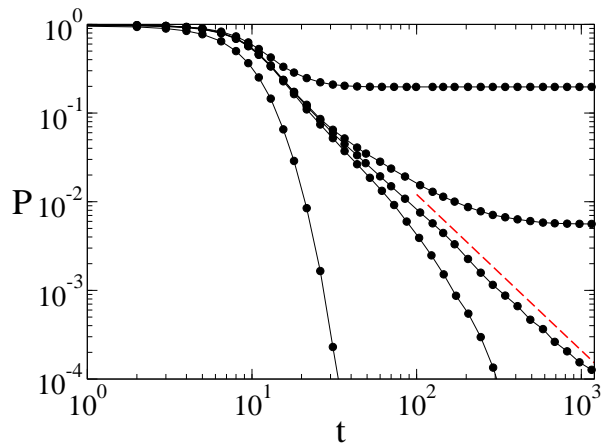}
\caption{Spreading experiments: probability $P(t)$ that the system is
  still alive at time $t$ in the Bilinguals model on a fully connected
  network, obtained from the same spreading experiments and parameters
  ($v=0$, $N=10^5$) as described in  Fig.~\ref{P-AS} for the
  Abrams-Strogatz model. The curves correspond to volatilities
  $a=0.600, 0.618, 0.620, 0.622$ and $0.700,$ (top to bottom).  $P(t)$
  decays as $t^{-\delta}$ at the transition point $0.620$ (close to
  the theoretical value $a_{fc} \simeq 0.63$), with $\delta \simeq
  1.76$, indicated by the dashed line.}
\label{P-MW}
\end{center}
\end{figure}

\subsubsection{Transition point for $v=0$}

In the symmetric case $v=0$ one can easily verify that the points
$(m=\pm 1, \sigma_0=0)$ in the $(m,\sigma_0)$ plane are two stationary
solutions of Eqs.~(\ref{dmdt5}) and (\ref{dsdt}).  But there is also a
third non-trivial stationary solution $(m=0, \sigma_0 = \sigma_0^*)$, where 
$\sigma_0^*$ satisfies
\begin{equation}
2 \sigma_0^* (1+\sigma_0^*)^a - (1-\sigma_0^*)^{(1+a)}=0.
\label{stat}
\end{equation}
By doing a small perturbation around $(0,\sigma_0^*)$ in the
$\sigma_0$ direction, one finds from Eq.~(\ref{dsdt}) that the point
$(0,\sigma_0^*)$ is stable for all values of $a$.   Instead, the
stability in the $m$ direction changes at some value $a_{fc}$ ($fc$ stands 
for fully connected).
Replacing $m$ by $\epsilon \ll 1 $ and $\sigma_0$ by $\sigma_0^*$ in
Eq.~(\ref{dmdt5}), one arrives to the following relation that
$\sigma_0^*$ and $a_{fc}$ hold when the stability changes
\begin{equation}
2\, a_{fc}\, \sigma_0^* (1+\sigma_0^*)^{(a_{fc}-1)} + (a_{fc}-1)
(1-\sigma_0^*)^{a_{fc}} = 0.
\label{stab}
\end{equation}
Combining Eqs.~(\ref{stat}) and (\ref{stab}), one arrives to the
following closed equation for $a_{fc}$
\begin{equation}
a_{fc}~\ln { \left( \frac{1-a_{fc}}{a_{fc}} \right) } = \ln  { \left(
  \frac{2a_{fc}-1}{1-a_{fc}} \right)},
\end{equation}
whose solution is $a_{fc} \simeq 0.63$.  Then, assuming that the
transition point does not depend on $v$ for FCN, as it happens in the
ASM, we find that the $(a,v)$ plane is divided into two regions.  In
the region $a<a_{fc}$, the stable solution is $(0,\sigma_0^*)$,
representing a stable mix of the three kinds of individuals, while in
the region $a>a_{fc}$, the stable solutions $(\pm 1,0)$ indicate the
ultimate dominance of one of the languages.  By performing spreading
experiments we estimated that the transition point for a network of
$N=10^5$ nodes is around $a = 0.62$ (see Fig.~\ref{P-MW}),
and we observed that this value approaches to the  analytical one
$a_{fc} \simeq 0.63$ as $N$ increases.  We have also checked numerically the
transition point when there is a bias ($v \neq 0$), that is, when the
two languages are not equivalent.  In this case we found a transition around 
$a = 0.675$ for a bias $v=-0.2$ and $N=10^5$ nodes, what represents
a small deviation from $a_{fc}$.  However this difference
is similar to the one found for the ASM in
Section~\ref{Fully-stability} with the same system size.  Therefore,
we assume that this discrepancy is again due to finite size effects and, in 
the thermodynamic limit, the transition should be at $a_{fc}$, for any value
of $v$.

We note that the transition point $a_{fc} \simeq 0.63$ is smaller than
the corresponding value $a_c \simeq 1.0$ for the ASM, thus
the region for coexistence is reduced in the BM.  This has a striking
consequence.  Suppose that there is population with individuals that
can use only one of two languages at a time, and it is characterized
by a volatility $a=0.8$, that allows the stable coexistence of the two
languages.  If now the behavior of the individuals is changed, so that
they can use both languages before they start using the
opposite language, the population looses the coexistence and finally
approaches to a state with the complete dominance of one language.  In
other words, within these models, bilinguals in use hinder language coexistence.

\subsubsection{AB Model: Neutral volatility and symmetric case}

For $a=1$ and $v=0$, Eqs.~(\ref{dmdt5}) and (\ref{dsdt}) are reduced
to
\begin{equation}
\frac{dm}{dt} = \frac{1}{2} \sigma_0 m,
\label{dmdt6}
\end{equation}
\begin{equation}
\frac{d \sigma_0}{dt} = \frac{1}{4} (1-m^2-4 \sigma_0 - \sigma_0^2).
\label{dsdt1}
\end{equation}
The three stationary solutions are $(m,\sigma_0) = (-1, 0)$; $(1,0)$
and $(0,\sqrt{5}-2)$.  Given that the above equations are difficult to
integrate analytically, we try an a approximate solution by assuming
that the density of bilinguals is proportional to the interface
density $\rho$, something observed in our simulations, and already
found in \cite{Castello-06} for the \emph{AB-model} (equivalent to the
BM in the case $v=0$ and $a=1$). Bilinguals are at the interface
between monolinguals, for all the networks studied.  Then we write
$\sigma_0 \simeq \alpha \rho$, where $\alpha$ is a constant and 
$\rho = 2 \sigma_- \sigma_+ + 2
\sigma_0 (\sigma_-+\sigma_+) = \frac{1}{2} \left[ (1-\sigma_0)^2 - m^2
  \right] + 2 \sigma_0 (1-\sigma_0)$, from where we obtain that $m$
can be expressed in terms of $\sigma_0$ as $m^2=(1-\sigma_0)^2+4
\sigma_0 (1-\sigma_0) - 2 \sigma_0/\alpha$.  Replacing this expression
for $m^2$ into Eq.~(\ref{dsdt1}), we obtain the following equation for
$\sigma_0$
\begin{equation}
\frac{d \sigma_0}{dt} = \frac{\sigma_0}{2} (-3+\frac{1}{\alpha} +
\sigma_0).
\end{equation}
We have checked by numerical simulations that $\alpha > 1/3$, then the
solution of the above equation in the long time limit is $\sigma_0
\sim e^{(-3+1/\alpha)t/2}$.  Therefore, $\sigma_0$ and $|m|$ approach
to $0$ and $1$, respectively, and the system reaches full order
exponentially fast.

\bigskip

\subsection{Complex networks}

We now consider the model on complex networks. Following the same
approach as in Section \ref{AS-nets}, it is possible to write down a
set of nine coupled differential equations:
three for the  densities $\sigma_-$, $\sigma_0$ and $\sigma_+$ of node
states, and six for the densities $\rho_{--}$, $\rho_{-0}$,
$\rho_{+0}$, $\rho_{+-}$, $\rho_{+0}$ and $\rho_{++}$ of different
types of links.  However, due to the complexity of these equations, we have 
limited our study to the investigation of the stability regions through 
Monte Carlo simulations.  We found that in a degree-regular random
graph with mean degree $\mu=3$, the stability diagram is qualitatively
similar to the one in Fig.~\ref{stab-DR} for the ASM, where the coexistence
region corresponds to stationary states with a mix of the three types of 
speakers.  Also, the 
coexistence-dominance transition point for $v=0$ is at 
$a_{cn} \simeq 0.3$ ($cn$ stands for complex networks).  For
$v=-0.02$, a monostable region appears for $0.2 \lesssim a \lesssim
0.4$, while this region becomes wider for $v=-0.2$ ($0 \lesssim a
\lesssim 1.4$). We have also observed that the coexistence region
disappears for $|v|\geq0.2$.  Therefore, in the BM, the region for
coexistence also shrinks as the connectivity of the network decreases
(going from fully connected to complex networks with low degree), but on
top of that, there exists a shift of the critical value from $a_{fc}
\simeq 0.63$ (fully connected networks) to $a_{cn} \simeq 0.3$
(degree-regular random graphs). In summary, compared to the ASM, the
overall effect of the inclusion of bilingual agents is that of a large
reduction of the region of coexistence.

\section{Square lattices}
\label{Square}

Dynamical properties of the ASM and BM in square lattices can be
explored for different initial conditions, system sizes, and values of
the prestige and volatility parameters, through a simulation applet
available online \cite{languageApplet}. It turns out that the behavior
of these models in square lattices is very different to their behavior
in fully connected or complex networks.  On the one hand, the mean
distance between two sites in the lattice grows linearly with the
length of the lattice side $L$, thus a spin only ``feels'' the spins
that are in its
near neighborhood, and therefore the mean-field approach that works
well in fully connected networks gives poor results in lattices.  On
the other hand, correlations between second, third and higher order
nearest-neighbors are important in lattices, what causes the formation
of same-spin domains, unlike in random networks where correlations to
second nearest-neighbors are already negligible.  Thus, pair
approximation does not provide a good enough description of the
dynamics in lattices either, and one is forced to implement higher
order approximations (triplets, quadruplets, etc), that lead to a
coupled system of many equations, impossible to solve analytically.
Due to the fact that the mean-field and pair approximations, that use
global quantities such as the magnetization and the density of
opposite-state links to describe the system, do not give good results
in lattices, we follow here a different approach to obtain a
macroscopic description. This approach, also developed in
\cite{Vazquez-08c} for general nonequilibrium spin models consists in
deriving a macroscopic equation for the evolution of a continuous
space dependent spin field. Within this approach it is possible to
describe coarsening processes, that is, processes of growth of local
linguistic domains caused by the motion of linguistic boundaries
(interface motion). In particular, one can explain whether the system
orders or not, or if the ordering is curvature driven (interface
motion due to surface tension reduction) or noise driven (without
surface tension).

We focus here on the ASM, but this macroscopic description can also be
applied for systems with three states, as the BM (see
\cite{Dall'Asta-08}).  Given that neighboring spins tend to be aligned
-due to the ferromagnetic nature of the interactions-, and also
correlations between spins reinforce the alignment between far
neighbors, the dynamics is characterized by the formation of same-spin
domains.  Starting from a well-mixed system with up and down spins
randomly distributed over the lattice, after a small transient, if we
look at the lattice from far we see domains growing and shrinking
slowly with time, and we can interpret this dynamics at the
coarse-grained level as the evolution of a continuous \emph{spin
  field} $\phi$ over space and time.  Then, we define by $\phi_{\bf
  r}(t)$ the spin field at site ${\bf r}$ at time $t$, which is a
continuous representation of the spin at that site ($-1<\phi<1$), also
interpreted as the average value of the spin over many realizations of
the dynamics.  Thus, we assume that there are $\Omega$ spin particles
at each site of the lattice, and we replace $\phi_{\bf r}(t)$ by the
average spin value $\phi_{\bf r}(t) \to \frac{1}{\Omega} \sum_{j=1}^\Omega S_{\bf
  r}^j$, where $S_{\bf  r}^j$ is the spin of the $j$-th particle
inside site ${\bf r}$.  Within this formulation, the dynamics is the
following.  In a time step of length $\delta t = 1/\Omega$, a site
${\bf r}$ and a particle from that site are chosen at random.  The
probability that the chosen particle has spin $s=\pm 1$ is equal to
the fraction of $\pm$ spins in that site $(1 \pm \phi_{\bf r})/2$.
Then the spin flips with probability
\begin{equation}
P(s  \to -s) = \frac{1}{2} (1- s v) \left(\frac{1 - s \psi_{\bf  r}}
{2}\right)^a,
\label{Ps-s2}
\end{equation}
where $\psi_{\bf r} \to \frac{1}{4} \sum_{{\bf r'/r}}\phi_{\bf r'}(t)$
is the average neighboring field of site ${\bf r}$, and the sum is
over the $4$ first nearest-neighbors sites ${\bf r'}$ of  site  ${\bf
  r}$.  If the flip happens, $\phi_{\bf r}$ changes by $-2 s/\Omega$,
thus its average change in time is given by the rate equation
\begin{equation}
\frac{\partial \phi_{\bf r}(t)}{\partial t} = \left[1-\phi_{\bf
    r}(t)\right] P(- \to +) - \left[1+\phi_{\bf r}(t)\right] P(+ \to -),
\label{dphi-dt}
\end{equation}
where the first (second) term corresponds to a $- \to +$ ($+ \to -$)
flip event.  In order to obtain a closed equation for $\phi$ (see
\ref{Eq-phi} for details), we substitute the expression for the
transition probabilities Eq.~(\ref{Ps-s2}) into Eq.~(\ref{dphi-dt}),
we then expand around $\psi_{\bf r}=0$, and replace the neighboring
field $\psi_{\bf r}$ by  $\phi_{\bf r} + \Delta \phi_{\bf r}$, where
$\Delta$ is defined as the standard Laplacian operator  $\Delta
\phi_{\bf r} \equiv  \frac{1}{4} \sum_{\bf r'/r} \left( \phi_{\bf
  r'}-\phi_{\bf r} \right) =\psi_{\bf r}-\phi_{\bf r}$.  Keeping the
expansion up to first order in $\Delta \phi_{\bf r}$, results in the
following equation for the spin field
\begin{eqnarray}
\frac{\partial \phi_{\bf r}(t)}{\partial t}  &=& 2^{-a}
\left(1-\phi_{\bf r}^2\right) \Biggl[ v+(a-1)\phi_{\bf r}+\frac{v}{2}
  (a-1)(a-2)\phi_{\bf r}^2 \nonumber \\ &+& 
\frac{1}{6} (a-1)(a-2)(a-3) \phi_{\bf r}^3 \Biggr]  \nonumber \\ 
&+& 2^{-a} a \left[ 1 + v (a-2) \phi_{\bf r} +
  \frac{1}{2} (a-1)(a-4) \phi_{\bf r}^2 \right] \Delta \phi_{\bf r}.
\label{dphi-dt2}
\end{eqnarray}
Equation~({\ref{dphi-dt2}) can be written in the form of a time
  dependent Ginzburg-Landau equation
\begin{equation}
\frac{\partial \phi_{\bf r}(t)}{\partial t} = D(\phi_{\bf r}) \Delta
\phi_{\bf r} - \frac{\partial V_{a,v}(\phi_{\bf r})}{\partial
  \phi_{\bf r}},
\end{equation}
with diffusion coefficient
\begin{eqnarray}
D(\phi_{\bf r}) \equiv 2^{-a} a \left[ 1 + v (a-2) \phi_{\bf r} +
  \frac{1}{2} (a-1)(a-4) \phi_{\bf r}^2 \right] 
\end{eqnarray}
and potential
\begin{eqnarray}
V_{a,v}(\phi_{\bf r}) &\equiv& 2^{-a} \Biggl\{-v \phi_{\bf r} -
\frac{1}{2}(a-1) \phi_{\bf r}^2 + \frac{v}{6}  \left[ 2-(a-1)(a-2)
  \right] \phi_{\bf r}^3 \nonumber \\  &+& \frac{1}{24} (a-1)
\left[6-(a-2)(a-3) \right] \phi_{\bf r}^4 +\frac{v}{10}
(a-1)(a-2)\phi_{\bf r}^5 \nonumber \\  &+& \frac{1}{36}(a-1)(a-2)(a-3)
\phi_{\bf r}^6 \Biggr\},
\end{eqnarray}
which is analogous to the potential for the global magnetization $m$
in the fully connected network case (Fig.~\ref{V-av}).  As we already
discussed in section \ref{Fully}, for the asymmetric case $v \not= 0$
the ordering dynamics is strongly determined by $v$.  When $a>1$,
$V_{a,v}$ has the shape of a double-well potential with minima at
$\phi = \pm 1$, and with a well deeper than the other, thus the system
is quickly driven by the bias towards the lowest minimum, reaching
full order in a rather short time.  For $a<1$ there is a minimum at
$|\phi|<1$, thus the system relaxes to a partially ordered state of
language coexistence composed by a well mixed population with
different proportions of speakers of the two languages.

\begin{figure}[t]
\begin{center}
 \includegraphics[width=0.6\textwidth]{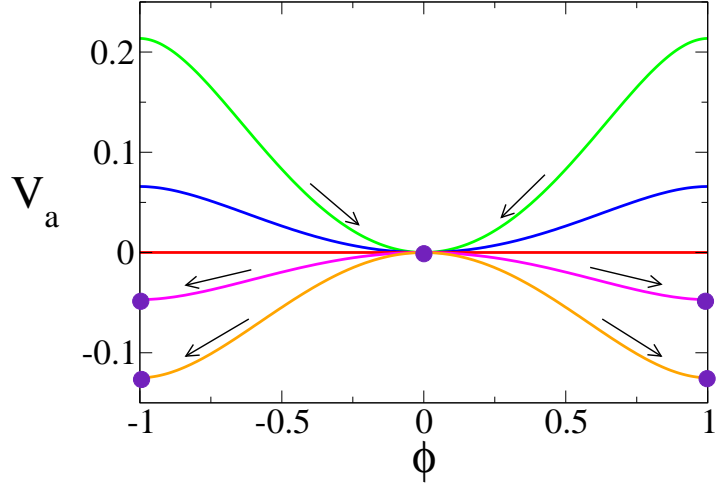}
 \caption{Ginzburg-Landau potential Eq.~(\ref{Va}) for the symmetric
   case $v=0$ of the Abrams-Strogatz model, with volatility values
   $a=0.5, 0.8, 1.0, 1.2$ and $2.0$ (from top to bottom).   For
   $a=0.5$ and $0.8$ the system relaxes to an active state with the
   same fraction of up and down spins uniformly distributed over the
   space, corresponding to the minimum of the potential at $\phi=0$,
   while for $a=1.2$ and $2.0$ it reaches full order, described by the
   field $|\phi|=1$.}
 \label{V-a}
\end{center}
\end{figure}

Specially interesting is the analysis of the symmetric case $v=0$, for
which the potential is (see Fig.~\ref{V-a})
\begin{eqnarray}
V_a(\phi_{\bf r}) = 2^{-a} (a-1) \biggl\{-\frac{\phi_{\bf r}^2}{2}+
\left[6-(a-2)(a-3)\right] \frac{\phi_{\bf
    r}^4}{24}+(a-2)(a-3)\frac{\phi_{\bf r}^6}{36} \biggr\}. \nonumber
\\
\label{Va}
\end{eqnarray}
In this bias-free case, when $a<1$ the minimum is at $\phi=0$, thus
the average magnetization in a small region around a given point ${\bf
  r}$ is zero, indicating that the system remains disordered (language
coexistence). This can be seen in Fig.~\ref{lattice}(b), where we show
a snapshot of the lattice for the model with $v=0$ and $a=0.5$, after
it has reached a stationary configuration.   For $a > 1$ the potential
has two wells with minima at $\phi=\pm 1$, but with the same depth,
thus there is no preference for any of the two states, and the system
orders in either of the language dominance states by spontaneous
symmetry breaking.  The order-disorder nonequilibrium transition at
$a=1$ is reminiscent of the well known Ising model transition, but
with the volatility parameter $a$ playing the role of temperature:
high volatility $a<1$ corresponds to the high temperature paramagnetic
phase and low volatility to the low temperature phase. An important
difference is that the transition is here first order, since the low
volatility stable sates $\phi=\pm 1$ appear discontinuously at
$a=1$. In addition, while in the low temperature phase of the Ising
model, spins flip in the bulk of ordered domains by thermal
fluctuations, here, spin flips in the low volatility regime only occur
at the interfaces (domain boundaries).

Complete ordering for $a > 1$ is achieved through domain coarsening
driven by surface tension \cite{Gunton_1983}.  That is, as the system
evolves, same-spin domains are formed, small domains tend to shrink
and disappear while large domains tend to grow.
Figure~\ref{lattice}(d) shows a snapshot of the lattice for the
evolution of the model with $a=2$.  We observe that domains have
rounded boundaries given that the dynamics tends to reduce their
curvature, leading to an average domain length that grows with time as
$l \sim t^{1/2}$ \cite{Castello-06,Dall'Asta-08}.  For the special
case $a=1$ (voter model) the potential is $V_a=0$, there is still
coarsening but without surface tension, meaning that domain boundaries
are driven by noise, as seen in Fig.~\ref{lattice}(c).  As a
consequence of this, the average length of domains grows very slowly
with time, as $l \sim \ln t$
\cite{Krapivsky-92,Frachebourg-96,Redner-01}.

\begin{figure}[t]
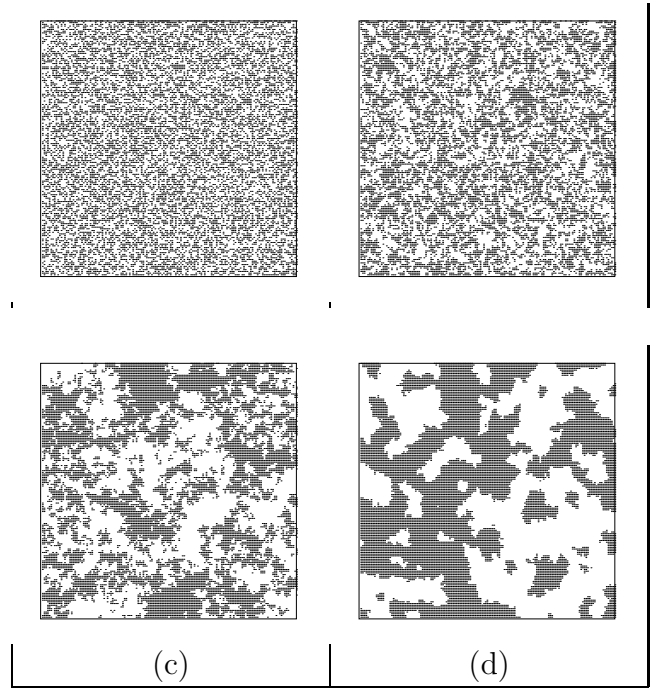

\begin{center}
\begin{tabular}{|c|c|}
\hline \includegraphics[width=1.5in, bb=70 70 550 550]{figure10a.eps}
& \includegraphics[width=1.5in, bb=70 70 550 550]{figure10b.eps} \\(a)
& (b) \\ \hline \includegraphics[width=1.5in, bb=70 70 550
  550]{figure10c.eps} & \includegraphics[width=1.5in, bb=70 70 550
  550]{figure10d.eps} \\(c) & (d) \\ \hline
\end{tabular}
 \caption{Snapshots of the Abrams-strogatz model with bias $v=0$ on a
   $128 \times 128$ square lattice and three values of volatility,
   $a=0.5$ (b), $a=1.0$ (c) and $a=2.0$ (d).  (a) Initial state: each
   site is occupied with a spin $+1$ or $-1$ with the same probability
   $1/2$. (b) The system reaches an active disordered stationary
   state, with a global magnetization that fluctuates around zero.
   (c) The system displays coarsening driven by noise, characterized
   by domains with
   noisy boundaries.  (d) There is also coarsening but driven by
   surface tension, generating domains with more rounded boundaries.}
\label{lattice}
\end{center}
\end{figure}

In order to compare the behavior of the language competition models
described before in fully connected and complex networks with their
behavior in square lattices we have numerically explored the stability
regions in the $(a,v)$ plane for the ASM and BM in square lattices.
The coexistence-dominance transition in the ASM for $v=0$ is at $a_c
\simeq 1.0$, as in fully connected and complex networks, whereas the
region for coexistence is found to be much more narrow than the ones
observed in complex networks with low degree, like the one depicted in
Fig.~\ref{stab-DR} for $\mu=3$.  Using the simulation applet 
\cite{languageApplet} one can
check that for a given value of $v \not=0$,  the disordered stationary
state that characterizes coexistence is harder to maintain in square
lattices than in random networks: in order to have an equivalent
situation, a smaller value of $a$ is needed in the former case.  

In the BM, apart from the narrowing of the coexistence region, we also
found that the transition point for $v=0$ is shifted to an even
smaller value of the volatility $a$ than in complex networks.  To see
this, in Fig.~\ref{r-MW} we show the time evolution of the inverse of
the average interface density  $\langle \rho \rangle$ 
\footnote{The $\langle...\rangle$ indicates average over independent 
realizations
  of the dynamics with different random initial conditions.} for
various values of $a$, on a square lattice of size $N=400^2$.     We
observe that $\langle \rho \rangle$ decays to zero for values of
$a>0.16$, indicating that the system orders (dominance phase), while
$\langle \rho \rangle$ approaches to a constant value larger than zero
for $a<0.16$, thus the system remains disordered (coexistence phase).
At the transition point $a_{sl} \simeq 0.16$ we have that $\langle
\rho \rangle \sim 1/\ln(t)$, indicating that the transition belongs to
the Generalized Voter class, a typical transition observed in spin
systems with two symmetric absorbing states
\cite{Al_Hammal_Chate-05,Vazquez-08c,Dornic2001,Droz2003}.  

The fact that $a_{sl} \simeq 0.16 $ is smaller than the corresponding
transition points $a_{fc}\simeq 0.63$ and $a_{cn} \simeq 0.3$,
together with the narrowing effect mentioned above, leads to the
result that the region for coexistence is largely reduced in square
lattices,  compared to fully connected and complex networks.

\begin{figure}[t]
\begin{center}
\includegraphics[width=0.6\textwidth]{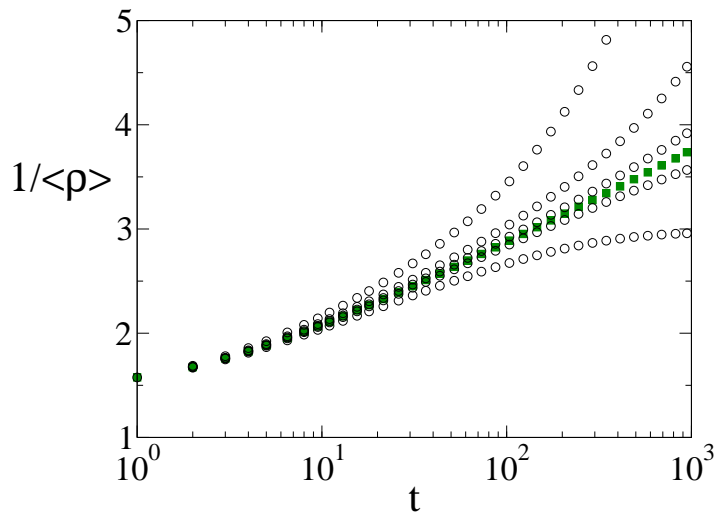}
\caption{Inverse of the average interface density  $\langle \rho
  \rangle$ vs time, on a log-linear scale, for the Bilinguals
  model. From top to bottom: $a=0.30$, $0.20$, $0.17$, $0.16$, $0.15$
  and $0.10$.  Averages were done over $10^3$ independent realizations
  on a square lattice of side $L=400$.  $\langle \rho \rangle$ decays
  as  $1/\ln(t)$ at the transition point $a_{sl} \simeq 0.16$
  (solid squares),  corresponding to the behavior of a Generalized voter
  transition in two dimensions.}
 \label{r-MW}
\end{center}
\end{figure}

\section{Summary and conclusions}
\label{Summary}

We have discussed the order-disorder transitions that occur in the
volatility-prestige parameter space of two related models of language
competition dynamics, the Abrams-Strogatz and its extension to account
for bilingualism: the Bilinguals Model. We have analyzed their
microscopic dynamics on fully connected, complex random networks and
two-dimensional square lattices and constructed macroscopic
descriptions of these dynamics accounting for the observed
transitions. At a general level, we have found that both models share
the same qualitative behavior, showing a transition from coexistence
to dominance of one of the languages at a critical value of the
volatility  parameter $a_c$.  The fact that agents are highly volatile
($a<a_c$), i.e, loosely attached to the language they are currently
using, leads to the enhancement of language coexistence. On the
contrary, in a low volatility regime ($a>a_c$), the final state is one
of dominance/extinction.

A more detailed comparison of both models shows important differences:
In the mean field description for fully connected networks, and for
the ASM, a scenario of coexistence is obtained for $a < 1$.  This is
independent of the relative prestige of the languages, $v$. However,
the stationary fraction of agents in the more prestigious language
increases with a higher prestige.  But when bilingual agents are
introduced (BM), the scenario of coexistence becomes the parameter
space area corresponding to $a<0.63$. That is, the area of coexistence
is reduced: agents with a higher level of volatility (smaller a) are
needed in order to obtain a coexistence regime. Within this current
framework, allowing the agents to use two languages at the same time,
which is reasonable from a sociolinguistic point of view, has the
effect of making language coexistence more difficult to achieve, in
the sense that coexistence occurs for a smaller range of parameters.

Network topology and local effects have been addressed through pair
approximations for degree uncorrelated networks. For the ASM on
degree-regular random networks we find that the decrease of the
network connectivity leads to a reduction, in the parameter space ($a,
v$), of the area of language coexistence and the area of bistable
dominance, while monostable dominance regions appear, in which only
the state of dominance of the more prestigious language is stable.  To
gain intuition on this result, we first notice that in the fully
connected network, the area of coexistence ($a<1$) corresponds to a
situation in which the majority of the agents use the more prestigious
language. The fact that all agents are interconnected, translates to a
situation in which users of the less prestigious language (minority)
are in contact with every other agent in the network. In this
situation, high volatility (agents switching their language use
easily) is effective in order to achieve a steady state situation with
individuals continuously changing the use of their language and making
coexistence possible.  In contrast, when considering a degree-regular
random network, that is, when limiting the number of neighbors in a
society, the existence of bias ($v\neq0$) opens the possibility for
agents in the majority language to be placed in domains without
contact with the minority language. For a region of the parameter
space where there is coexistence in a fully connected network, these
domains can grow in size in a random network until they occupy the
entire system. This gives rise to the monostable region of dominance
of the more prestigious language found in complex networks with low
connectivity. Compared to the fully connected case, a higher
volatility is needed in order to overcome this topological effect,
leading to a reduction of the area of coexistence.  In two dimensional
square lattices the coexistence is shown to be even more difficult to
achieve, probably due to the fact that correlations with second
neighbors make the coarsening process of formation and growth of
domains easier.  The  macroscopic field description introduced for
square lattices accounts for the different coarsening processes
observed for large and small volatility.

The network effects described above for the ASM are also qualitatively
valid for the BM. However, the reduction of the area of language
coexistence is more important when considering bilingual agents. We
find a shift of the critical value with the topology: $a_{fc} \simeq
0.63$ in fully connected networks, $a_{cn} \simeq 0.3$ in complex
uncorrelated networks, and $a_{sl} \simeq 0.16$ in two dimensional
square lattices.

In summary, building upon previous works on language competition
\cite{Abrams-03, Wang-05_TRENDS_Ecology, Minett-08}, we have studied
numerically and by analytical macroscopic descriptions, two
microscopic models for the dynamics of language competition.  We have
analyzed the role of bilingual agents and social network structure in
the order-disorder transitions occurring for different values of the
two parameters of the models: the relative prestige of the languages
and the volatility of the agents.  We have found that the scenario of
coexistence of the two languages is reduced when bilingual agents are
considered.  This reduction also depends on the social structure, with
the region of coexistence shrinking when the connectivity of the
network decreases.

We acknowledge  support from project FISICOS (FIS2009-60327) of MEC
and  FEDER, and from NEST-Complexity project PATRES (043268).



\bigskip

\begin{thebibliography}{10}

\bibitem{Castellano-09-a}
C.~Castellano, S.~Fortunato, and V.~Loreto.
\newblock Statistical physics of social dynamics.
\newblock {\em Reviews of Modern Physics}, 81:591, 2009.

\bibitem{SanMiguel-05_CISE}
M.~San~Miguel, V.M. Egu\'{\i}luz, R.~Toral, and K.~Klemm.
\newblock Binary and multivariate stochastic models of consensus formation.
\newblock {\em Computer in Science and Engineering}, 7:67--73, 2005.

\bibitem{Appel_1987}
R.~Appel and P.~Muysken.
\newblock {\em Language Contact and Bilingualism}.
\newblock Edward Arnold Publishers, London, 1987.

\bibitem{Mira-05}
J.~Mira and A.~Paredes.
\newblock Interlinguistic similarity and language death dynamics.
\newblock {\em Europhysics Letters}, 69:1031--1034, 2005.

\bibitem{Abrams-03}
D.~M. Abrams and S.~H. Strogatz.
\newblock Modelling the dynamics of language death.
\newblock {\em Nature}, 424:900, 2003.

\bibitem{Crystal-00}
D.~Crystal.
\newblock {\em Language death}.
\newblock Cambridge University Press, Cambridge, 2000.

\bibitem{Stauffer-07}
D.~Stauffer, X.~Castell\'o, V.~M. Egu\'iluz, and M.~San~Miguel.
\newblock Microscopic abrams-strogatz model of language competition.
\newblock {\em Physica A}, 374:835--842, 2007.

\bibitem{Patriarca_2008}
M.~Patriarca and E.~Heinsalu.
\newblock Influence of geography on language competition.
\newblock {\em Physica A}, 388:174--186, 2009.

\bibitem{Pinasco-06}
J.P. Pinasco and L.~Romanelli.
\newblock Coexistence of languages is possible.
\newblock {\em Physica A}, 361(1):355--360, February 2006.

\bibitem{Schulze-05}
C.~Schulze and D.~Stauffer.
\newblock Monte carlo simulation of the rise and the fall of languages.
\newblock {\em International Journal of Modern Physics C}, 16(5):781--787,
  2005.

\bibitem{Vivane-06(1)}
Viviane~M. de~Oliveira, M.A.F. Gomes, and I.R. Tsang.
\newblock Theoretical model for the evolution of the linguistic diversity.
\newblock {\em Physica A}, 361(1):361--370, February 2006.

\bibitem{Ligget_1985}
T.~M. Liggett.
\newblock {\em Interacting Particle Systems}.
\newblock New York: Springer, 1985.

\bibitem{Marro_Dickman_1999}
J.~Marro and R.~Dickman.
\newblock {\em Nonequilirium Phase Transitions in Lattice Models}.
\newblock Cambridge University Press: Cambridge, U.K., 1999.

\bibitem{Wang-05_TRENDS_Ecology}
William S-Y. Wang and James~W. Minett.
\newblock The invasion of language: emergence, change and death.
\newblock {\em Trends in Ecology and Evolution}, 20:263--269, 2005.

\bibitem{Minett-08}
J.~W. Minett and W.~S.-Y. Wang.
\newblock Modelling endangered languages: The effects of bilingualism and
  social structure.
\newblock {\em Lingua}, 118:19--45, 2008.

\bibitem{Castello-06}
X.~Castell\'o, V.~M. Egu\'{\i}luz, and M.~San~Miguel.
\newblock Ordering dynamics with two non-excluding options: bilingualism in
  language competition.
\newblock {\em New Journal of Physics}, 8:308--322, 2006.

\bibitem{Castello-07}
X.~Castell\'o, R.~Toivonen, V.~M. Egu\'{\i}luz, J.~Saram\"aki, K.~Kaski, and
  M.~San~Miguel.
\newblock Anomalous lifetime distributions and topological traps in ordering
  dynamics.
\newblock {\em Europhysics Letters}, 79:66006 (1--6), 2007.

\bibitem{Toivonen-08}
R.~Toivonen, X.~Castell\'o, V.~M. Egu\'{\i}luz, J.~Saram\"aki, K.~Kaski, and
  M.~San~Miguel.
\newblock Broad lifetime distributions for ordering dynamics in complex
  networks.
\newblock {\em Physical Review E}, 79:016109, 2008.

\bibitem{Dall'Asta-07}
L.~Dall'Asta and C.~Castellano.
\newblock Effective surface-tension in the noise-reduced voter model.
\newblock {\em Europhysics Letters}, 77:60005, 2007.

\bibitem{Stark-08}
H.~U. Stark, C.~J. Tessone, and F.~Schweitzer.
\newblock Slower is faster: Fostering consensus formation by heterogeneous
  inertia.
\newblock {\em Advances in Complex Systems}, 11:551, 2008.

\bibitem{blythe2009gmc}
R.~A. Blythe.
\newblock {Generic modes of consensus formation in stochastic language
  dynamics}.
\newblock {\em Journal of Statistical Mechanics}, P02059, 2009.

\bibitem{Vazquez-03}
F~Vazquez, P~L Krapivsky, and S~Redner.
\newblock Constrained opinion dynamics: freezing and slow evolution.
\newblock {\em J. Phys. A: Math. Gen.}, 36:L61--L68, 2003.

\bibitem{Vazquez-04}
F~Vazquez, , and S~Redner.
\newblock Ultimate fate of constrained voters.
\newblock {\em J. Phys. A: Math. Gen.}, 37:8479--8494, 2004.

\bibitem{Dall'Asta-08}
L.~Dall'Asta and T.~Galla.
\newblock Algebraic coarsening in voter models with intermediate states.
\newblock {\em J. Phys. A: Math. Theor.}, 41:435003, 2008.

\bibitem{Al_Hammal_Chate-05}
O.~Al~Hammal, H.~Chat\'e, I~Dornic, and M.~A. Mu\~{n}oz.
\newblock Langevin description of critical phenomena with two symmetric
  absorbing states.
\newblock {\em Physical Review Letters}, 94:230601, 2005.

\bibitem{Vazquez-08c}
F.~Vazquez and C.~L\'opez.
\newblock Systems with two symmetric absorbing states: relating the microscopic
  dynamics with the macroscopic behavior.
\newblock {\em Phys. Rev. E}, 78(061127), 2008.

\bibitem{Vazquez-08a}
F.~Vazquez and V.~M. Eguiluz.
\newblock Analytical solution of the voter model on uncorrelated networks.
\newblock {\em New Journal of Physics}, 10(063011), 2008.

\bibitem{ER}
P.~Erd\H{o}s and A.~R\'enyi.
\newblock On the evolution of random graphs.
\newblock {\em Publ.Math. (Debrecen)}, 6:290--297, 1959.

\bibitem{Barabasi_1999}
A.-L. Barab\'asi and R.~Albert.
\newblock Emergence of scaling in random networks.
\newblock {\em Science}, 286:509, 1999.

\bibitem{Castellano-09-b}
E.~Pugliese and C.~Castellano.
\newblock Heterogeneous pair approximation for voter models on networks.
\newblock {\em Europhysics Letters}, 88:58004 (1--6), 2009.

\bibitem{languageApplet}
X.~Castell\'o.
\newblock Dynamics of language competition.
\newblock {\em From IFISC website:
  http://www.ifisc.uib-csic.es/research/complex/APPLET\_LANGDYN.html}, 2007.

\bibitem{Gunton_1983}
J.~D. Gunton, M.~San~Miguel, and P.~Sahni.
\newblock {\em Phase Transitions and Critical Phenomena}, volume~8, chapter The
  dynamics of first order phase transitions, pages 269--446.
\newblock Academic Press, London, 1983.

\bibitem{Krapivsky-92}
P.~L. Krapivsky.
\newblock Kinetics of monomer-monomer surface catalytic reactions.
\newblock {\em Phys. Rev. A}, 45(2):1067--1072, Jan 1992.

\bibitem{Frachebourg-96}
L.~Frachebourg and P.~L. Krapivsky.
\newblock Exact results for kinetics of catalytic reactions.
\newblock {\em Physical Review E}, 53:R3009, 1996.

\bibitem{Redner-01}
Sidney Redner.
\newblock {\em A Guide to First-Passage Processes}.
\newblock Cambridge University Press, August 2001.

\bibitem{Dornic2001}
I.~Dornic, H.~Chat\'e, J.~Chave, and H.~Hinrichsen.
\newblock Critical coarsening without surface tension: The universality class
  of the voter model.
\newblock {\em Physical Review Letters}, 87:045701, 2001.

\bibitem{Droz2003}
M.~Droz, A.~L. Ferreira, and A.~Lipowski.
\newblock Splitting the voter potts model critical point.
\newblock {\em Physical Review E}, 67:056108, 2003.

\end{thebibliography}

\clearpage
\appendix

\section{Equation for the spin field $\phi_{\bf r}$}
\label{Eq-phi}

In this section we shall derive an equation for the spin field
$\phi_{\bf r}$.  We start by substituting the expression for the
transition probabilities Eq.~(\ref{Ps-s2}) into Eq.~(\ref{dphi-dt})
and by writing it in the more convenient form
\begin{eqnarray}
\frac{\partial \phi}{\partial t} = \frac{(1+v)}{2^{a+1}} (1-\phi)
(1+\psi) (1+\psi)^{a-1} - \frac{(1-v)}{2^{a+1}} (1+\phi) (1-\psi)
(1-\psi)^{a-1}, \nonumber \\ 
\label{dphi-dt3}
\end{eqnarray}
where $\phi$ and $\psi$ are abbreviated forms of $\phi_{\bf r}$ and
$\psi_{\bf r}$ respectively.  We now replace the neighboring field
$\psi$ in the terms $(1+\psi)$ and $(1-\psi)$ of Eq.~(\ref{dphi-dt3})
by $\psi \equiv \phi + \Delta \phi$, where $\Delta$ is defined as the
standard Laplacian operator $\Delta \phi_{\bf r} \equiv  \frac{1}{4}
\sum_{\bf r'/r} \left( \phi_{\bf r'}-\phi_{\bf r} \right) =\psi_{\bf
  r}-\phi_{\bf r}$, and obtain
\begin{eqnarray}
\frac{\partial \phi}{\partial t} &=& 2^{-(a+1)} (1-\phi^2) \left[
  (1+v) (1+\psi)^{a-1} - (1-v) (1-\psi)^{a-1} \right] \\ &+&
2^{-(a+1)} \left[ (1+v) (1-\phi) (1+\psi)^{a-1} + (1-v) (1+\phi)
  (1-\psi)^{a-1} \right] \Delta \phi. \nonumber 
\label{dphi-dt4}
\end{eqnarray}
Because our idea is to obtain a Ginzburg-Landau equation with a
$\phi^6$-potential, the right hand side of Eq.~(\ref{dphi-dt4}) must
be proportional to $\phi^5$, and therefore we use the Taylor series
expansions around $\psi=0$
\begin{eqnarray*}
(1 \pm \psi)^{a-1} &=& 1 + (a-1) \psi + \frac{1}{2}(a-1)(a-2) \psi^2 +
  \frac{1}{6}(a-1)(a-2)(a-3) \psi^3 ~~\mbox{and} \\ (1-\psi)^{a-1} &=&
  1 - (a-1) \psi + \frac{1}{2}(a-1)(a-2) \psi^2 -
  \frac{1}{6}(a-1)(a-2)(a-3) \psi^3
\end{eqnarray*}
into Eq.~(\ref{dphi-dt4}), to obtain
\begin{eqnarray}
\frac{\partial \phi}{\partial t} &=& 2^{-a} (1-\phi^2) \left[ v+(a-1)
  \psi +\frac{v}{2} (a-1)(a-2) \psi^2 + \frac{1}{6} (a-1)(a-2)(a-3)
  \psi^3 \right] \nonumber \\ &+& 2^{-a} \Biggl\{ (1-v \phi) \left[
  1+\frac{1}{2} (a-1)(a-2) \psi^2 \right] \nonumber \\ &+& (v-\phi)
\left[ (a-1) \psi + \frac{1}{6} (a-1)(a-2)(a-3) \psi^3 \right]
\Biggr\} \Delta \phi.
\label{dphi-dt5}
\end{eqnarray}
We then replace $\psi$ by $\phi + \Delta \phi$ in
Eq.~({\ref{dphi-dt5}) and expand to first order in $\Delta \phi$,
  assuming that the field $\phi$ is smooth, so that $\Delta \phi \ll
  \phi$.  Finally, neglecting $\phi^3$ and higher order terms in the
  diffusion coefficient that multiplies the laplacian, we arrive to
  the expression for the spin field quoted in Eq.~(\ref{dphi-dt2}).

\end{document}